\documentclass[nonacm, acmtog]{acmart}

\usepackage[ruled]{algorithm2e} 

\SetAlFnt{\small}
\SetAlCapFnt{\small}
\SetAlCapNameFnt{\small}
\SetAlCapHSkip{0pt}

\let\oldnl\nl
\newcommand{\nonl}{\renewcommand{\nl}{\let\nl\oldnl}}

\newlength\inlen
\newcommand\myinput[1]{%
  \nonl\settowidth\inlen{\KwIn{}}%
  \setlength\hangindent{\inlen}%
  \hspace*{\inlen}#1\\}

\newlength\outlen

\usepackage{xcolor}
\definecolor{greenish}{RGB}{0,109,44}

\def \bezier {B{\'e}zier}

\begin{document}
\title{Robust Containment Queries over Collections of Rational Parametric Curves via Generalized Winding Numbers}

\author{Jacob Spainhour}
\orcid{0000-0001-8219-4360}
\affiliation{%
  \institution{University of Colorado Boulder}
  \country{USA}}
\email{Jacob.Spainhour@colorado.edu}

\author{David Gunderman}
\orcid{0000-0003-4154-0911}
\affiliation{%
  \institution{Indiana University School of Medicine}
  \country{USA}}
\email{dgunderm@purdue.edu}

\author{Kenneth Weiss}
\orcid{0000-0001-6649-8022}
\affiliation{%
  \institution{Lawrence Livermore National Laboratory}
  \country{USA}}
\email{kweiss@llnl.gov}

\renewcommand\shortauthors{J. Spainhour, D. Gunderman and K. Weiss}

\begin{abstract}
Point containment queries for regions bound by watertight geometric surfaces, i.e., closed and without self-intersections, can be evaluated straightforwardly with a number of well-studied algorithms. 
When this assumption on domain geometry is not met, such methods are either unusable, or prone to misclassifications that can lead to cascading errors in downstream applications. 
More robust point classification schemes based on generalized winding numbers have been proposed, as they are indifferent to these imperfections.
However, existing algorithms are limited to point clouds and collections of linear elements.
We extend this methodology to encompass more general curved shapes with an algorithm that evaluates the winding number scalar field over unstructured collections of rational parametric curves. 
In particular, we evaluate the winding number for each curve independently, making the derived containment query robust to how the curves are arranged.
We ensure geometric fidelity in our queries by treating each curve as equivalent to an adaptively constructed polyline that provably has the same generalized winding number at the point of interest.
Our algorithm is numerically stable for points that are arbitrarily close to the model, and explicitly treats points that are coincident with curves.
We demonstrate the improvements in computational performance granted by this method over conventional techniques as well as the robustness induced by its application.
\end{abstract}

%
%

%
%


\begin{teaserfigure}
\centering
  \includegraphics[width=0.9\linewidth]{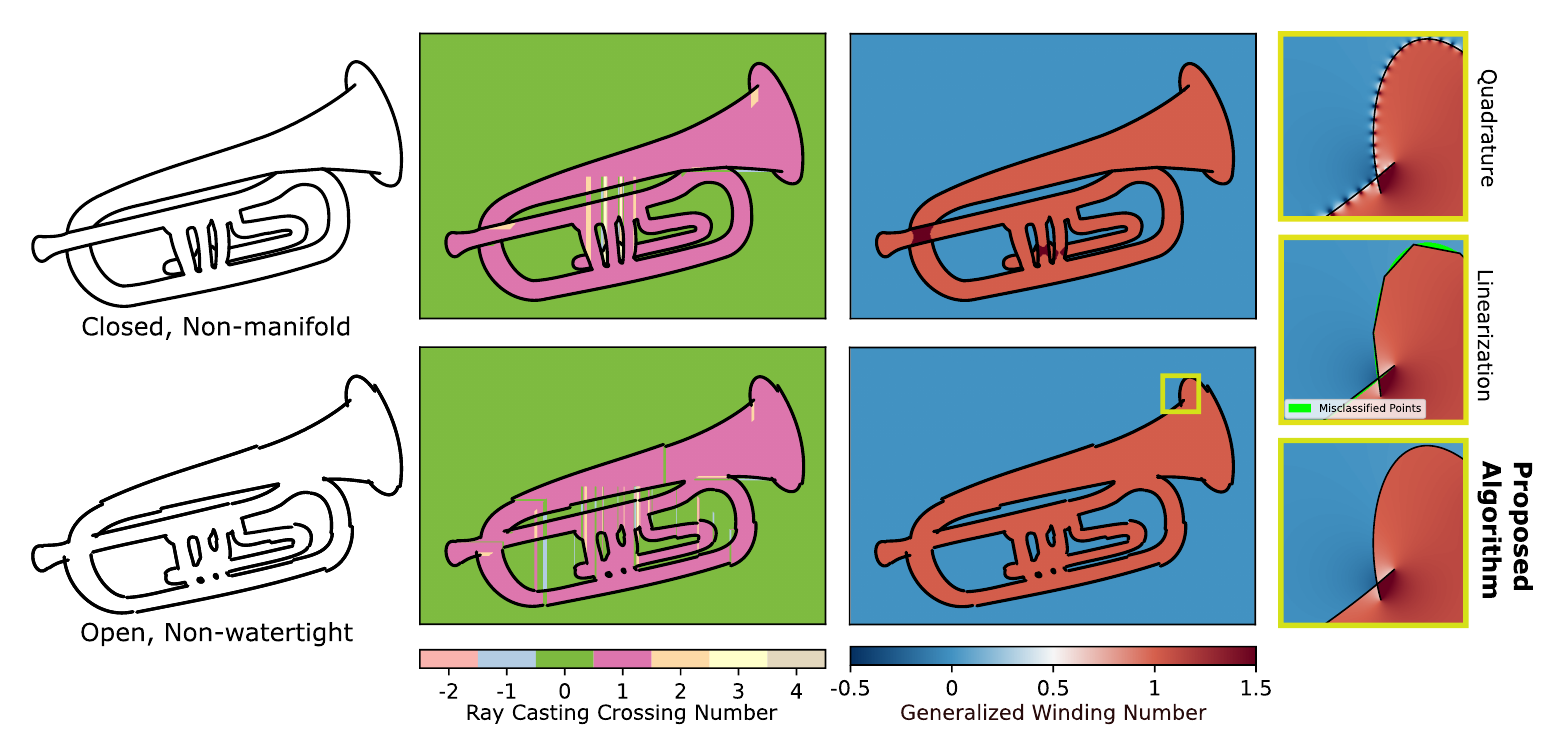}
\caption{The point containment query is an important predicate in many applications, yet their evaluation is limited by undesirable, but often unavoidable features of the bounding geometry such as non-manifold or non-watertight edges (left). 
This causes catastrophic errors for traditional approaches to deciding containment, in particular those that use ray casting (middle left). 
Because the field of generalized winding numbers degrades smoothly around geometric errors, they are well-suited for use in robust containment queries (middle right). 
In this paper, we present a framework for evaluating exact generalized winding numbers for arbitrary collections of curved objects, as well as a novel point-in-curved-polygon algorithm that facilitates their efficient calculation (right).}
\Description{Figure 1: A comparison between containment queries using traditional ray casting and generalized winding numbers is shown for a simple vector graphic. The ray casting method fails to correctly classify points near gaps, overlaps, or non-maniforld edges, while the field of generalized winding numbers is smooth around these artifacts. Three methods of computing generalized winding numbers are shown on a zoomed-in portion of the figure, including an unstable quadrature, an inaccurate linear approximation, and our proposed adaptive algorithm.}
\label{fig:graphical_abstract}
\end{teaserfigure}

\maketitle

\section{Introduction}\label{sec:Introduction}

In the fields of computer graphics, Computer Aided Geometric Design (CAGD) and Computer Aided Engineering (CAE), it is common for geometric objects to be expressed using a boundary representation (B-Rep).
Often, these B-Reps are defined via non-uniform rational B-spline (NURBS) curves in 2D and surfaces in 3D. 
Such objects are invaluable in the design of CAD models, as they allow for precise geometric control of the object's boundary~\cite{piegl-96-nurbs}.
However, it can still be desirable in many contexts, such as in animation or simulation, to treat the interior volume of these objects explicitly~\cite{mcadams-11-elasticity}.

Such a task motivates development of the \textit{containment query}, a geometric predicate that returns whether an arbitrary spatial location is contained within an arbitrary shape.
The rapid evaluation of containment queries is quite well understood for many classes of shapes, such as polygons and regions bounded by \bezier\ curves~\cite{nishita-90-raytracing, Carvalho-95-pointinpolyhedron, patrikalakis-10-interrogation}.
However, such methods assume the boundary is \textit{watertight}, having no gaps between connected components.
In reality, the vast majority of B-Reps are created by hand within some form of CAD software, leading to so-called ``messy'' CAD geometry. 

\begin{figure}[tbp]
    \centering
    \includegraphics[width=\linewidth]{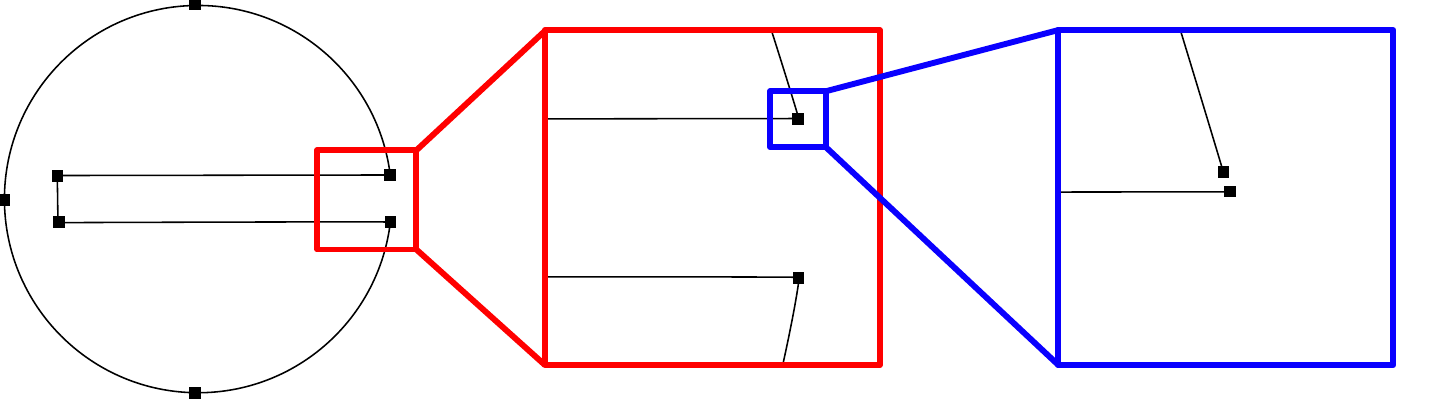}
    \caption{Geometric errors can be visually imperceptible, but can still cause a shape to have no topological interior.}
    \Description{Figure 2: We zoom in on a vertex of a simple 2D shape, which reveals that the model is unexpectedly non-watertight.}
    \label{fig:very_small_gaps}
\end{figure}

A straightforward source of these problems is human error during design or construction, but more subtle issues can be introduced through varying tolerances within and between different software tools.
This results in boundary models with human-imperceptible, but numerically significant gaps and overlaps between individual components of the boundary~\cite{marussig-17-isogeometricreview} (see Figure~\ref{fig:very_small_gaps}).
Even in situations prioritizing visibility over strict geometric fidelity, such as in computer graphics and animation, care must be taken to ensure that these errors are properly accounted for.
For example, the \emph{dynamic planar map} heuristic used by Adobe Illustrator's Live Paint Bucket tool allows users to set fills for closed regions bounded by intersecting collections of \bezier\ curves and exposes a global threshold parameter to connect gaps between curves in the collection~\cite{asente-07-live-paint}.

As an additional class of motivating examples, we consider applications in multiphysics simulation, for which containment classification errors can lead to cascading problems and incorrect results.
The mathematical underpinnings of containment within non-watertight shapes is dubious, which leads to a tenuous theoretical foundation for established methods that assume geometric continuity.
This is problematic, as errors are frequently inconsistent within local regions, with even arbitrarily close points receiving different classifications.
More practically, these errors can result in unexpected and fatal errors in simulation software, wasting computational resources and developer time~\cite{sederberg-08-watertight}.

At the same time, physical applications often prevent approximating curved geometry with piecewise linear segments. 
Although this would allow for treatment of the messy---but now linear---geometry with established methods, such an approximation can produce sizable errors in the relevant numerical method~\cite{sevilla-08-nurbsfem}.
As a concrete example, multimaterial Arbitrary Lagrangian-Eulerian (ALE) methods require additional preprocessing to initialize the volume fraction of each material in a computational cell~\cite{hirt-74-ale,barlow-ale}.
This, in turn, necessitates classifying each point in a quadrature rule along the background grid as inside or outside the volume bounded by the corresponding shape~\cite{weiss-16-shaping}.
It is critical that this identification takes into account the full curvature of the bounding geometry, as piecewise linear approximations can cause quadrature points near material boundaries to be misclassified, unpredictably compounding the error in the calculation.

Such geometric errors can be resolved with techniques in surface repair~\cite{mezentsev-99-surfacerepair,park-21-discretization,bischoff-05-mesh-repair}, which alter the B-Rep itself to an approximated surface that lacks these imperfections.
However, doing so by hand can be impractical for the typical number of imperfections in such a model, and automated procedures during geometry pre-processing can inadvertently remove important model features.
Instead, we require a containment query that is \textit{robust} to such artifacts, in the sense that an appropriate inside/outside classification is always returned, independent of size and frequency of these boundary errors.

To address these concerns, we present in this work three principal contributions:
\begin{itemize}
    \item We extend the theoretical framework of generalized winding numbers~\cite{Jacobson-13-winding} to the context of an unstructured collection of 2D rational parametric curves, from which one can derive a robust containment query that indicates whether an arbitrary point should be considered interior to non-watertight geometry.
    \item We address the issue of defining and evaluating generalized winding numbers for points that are coincident with the boundary.
    This discussion lies outside the scope of much of the existing literature, as such points require unique consideration in the context of curved geometry.
    \item We provide a novel algorithm for evaluating integer winding numbers with respect to curved geometry, which is the principal component of our technique for evaluating generalized winding numbers on such shapes. We demonstrate that this calculation is robust to non-watertight geometry and considerably more performant than state-of-the-art techniques. 
\end{itemize}
Altogether, these contributions allow for robust containment queries of objects implied by messy 2D CAD geometry, composed of collections of parametric curves (see Figure~\ref{fig:graphical_abstract}).
Our algorithm has been implemented in Axom, a BSD-licensed open-source library of Computer Science infrastructure for HPC applications~\cite{Axom_CS_infrastructure}.
\section{Background and Related Work}\label{sec:background}
\begin{figure}[tbp]
  \centering
  \includegraphics[width=\linewidth]{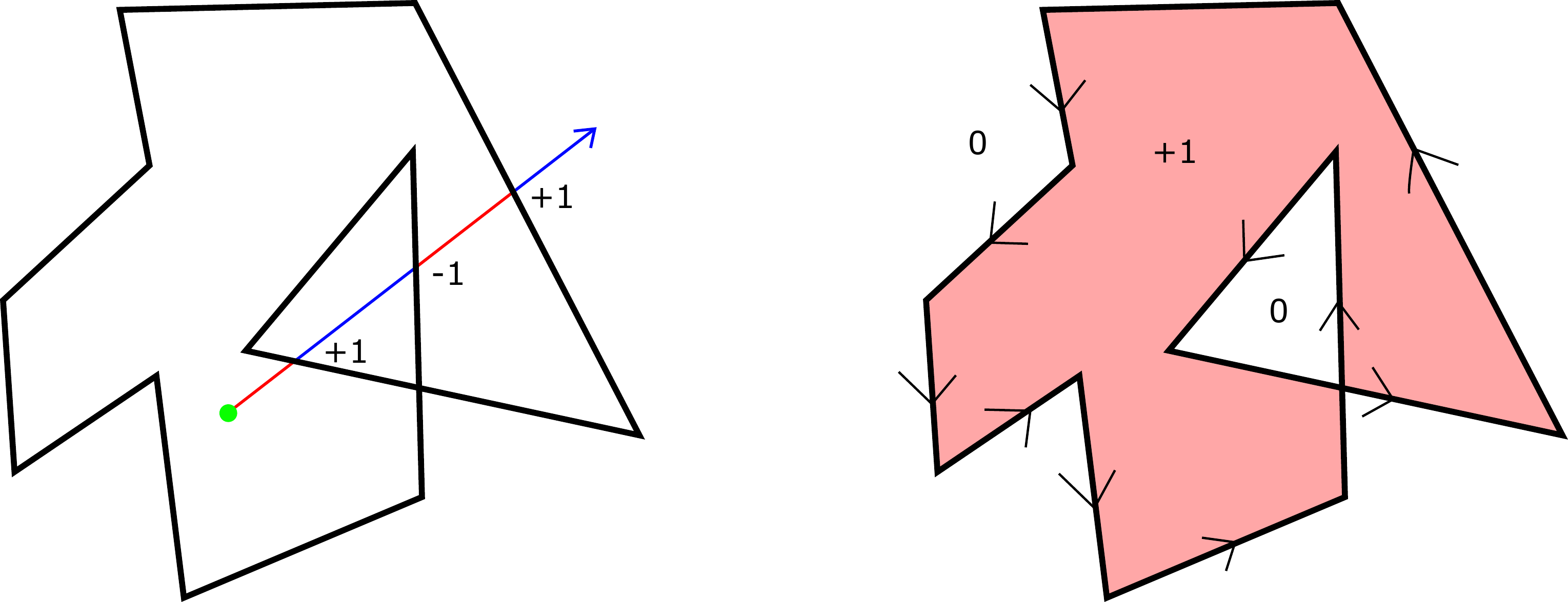}
  \caption{Given an arbitrary query point, containment can be determined with a ray casting algorithm 
           that counts (signed) \textit{intersections} between the polygon and a ray extending from the query (left), 
           or a winding number algorithm that counts \textit{revolutions} of the curve around the query (right).}
  \Description{Figure 3: Fully described in the caption.}
  \label{fig:ray_casting_vs_winding_number}
\end{figure}

Even in the ideal and watertight setting, methods of evaluating containment vary wildly depending on the numerical representation of the bounding geometry.
As simple examples, the operation can be performed trivially for axis-aligned quadrilaterals through the evaluation of inequalities, for spheres by computing a distance to the center, and for triangles using Barycentric coordinates.
More complicated procedures are necessary for point containment in arbitrary polygons~\cite{hains-94-strategies}.
These algorithms can generally be broken down into two categories (see Figure~\ref{fig:ray_casting_vs_winding_number}).

\textit{Ray Casting} algorithms determine containment by extending a ray from the specified query point out to infinity, and counting the number of times the ray intersects the domain boundary to produce a \textit{crossing number}~\cite{shimrat-62-positionofpoint}.
As a point moves along this ray, it will alternate between interior and exterior, and therefore an odd (even) crossing number indicates the point is interior (exterior).
Computing intersections between the given ray and the linear edges of a polygon can be done very efficiently, with many graphics processing units (GPUs) having dedicated ray-tracing cores that can perform the operation massively in parallel~\cite{zellmann-22-raytracing}.
However, these methods are sensitive to the specific ray that is extended, needing to account for special cases when the ray intersects a polygon vertex or entire edge, which can be unnecessarily more costly~\cite{edelsbrunner-90-simulation,sommariva-22-inrs}.
\textit{Winding number} algorithms are a class of algorithms that are more indifferent to these issues, and, as such, are the type from which we derive our proposed method.
In a winding number algorithm, we instead count the number of revolutions around the query point made by a particle traveling on the domain boundary~\cite{hormann-2001-pointinpolygon}.
This value can be considered a generalization of the crossing number, and can be used to determine containment with the same even-odd rule.
By virtue of not being dependent on any particular extended ray, these algorithms implicitly handle edge-cases that are problematic for ray casting.
This means even in the context of simple, watertight geometry, such algorithms can be considered a more robust approach to containment queries.

Containment of arbitrary points in more general curved shapes is a much more difficult problem, and has been studied extensively in the context of 2D vector graphics.
As an example, we refer to the process of \textit{linearization}, in which a curve is approximated by connected linear segments, an approach that is fairly common in both modern graphics engines and physics codes.
In the former, information about the targeted display can be leveraged to ensure that the approximating polyline is constructed with sufficiently high resolution such that it is indistinguishable to a viewer~\cite{kilgard-20-polarstroking}.
Combined with the fact that the polyline must be rendered with some predetermined thickness, it is therefore theoretically possible (although rarely enforced) for containment queries on linearized watertight geometries on a rendered display to have no \textit{observable} misclassifications.

On the other hand, this type of polyline approximation of curved geometry is known to cause severe and often unexpected errors in the modeling of physical systems through PDEs and finite element analysis~\cite{sevilla-08-nurbsfem}.
This is a particular problem in the case where points of interest are allowed to be placed arbitrarily close to the curve, as then no guarantee can be made as to how many refinements in the approximation are necessary to achieve perfect geometric fidelity.
In place of such a linearization, the use of NURBS to represent boundaries in a finite element analysis has been quite successful in reducing or even eliminating geometric error~\cite{sevilla-08-nurbsfem, sevilla-11-nurbsfem3d}.
As a result of this, we consider operating directly on curves without any approximation to be a necessary component of our proposed algorithm so as to preserve the accuracy of an underlying numerical method.

We similarly desire an algorithm that can reach this level of geometric fidelity for a very general class of curves.
For example, cubic \bezier\ curves are the highest order required by postscript and SVG formats, as the level of detail they afford is typically sufficient for design applications.
However, applications in computational multiphysics can necessitate an algorithm that works for \textit{rational} \bezier\ curves of \textit{arbitrary} order.
This is because the NURBS shapes that are used by more sophisticated modeling software can always be decomposed into a collection of (in principle arbitrary order) rational \bezier\ primitives through the process of \bezier\ extraction~\cite{piegl-96-nurbs,thomas-15-bezier-projection}.
However as we will see, the proposed algorithm is ultimately indifferent to the order and/or rationality of the curve in question.

Containment within regions defined by \bezier\ curves has also been studied in the context of 2D parametric trimming curves for 3D NURBS surface patches.
In this sub-problem, it must be determined if a point in the 2D parameter space of the 3D surface is contained within the (often, but not necessarily watertight) trimming curves.
Because this and other contemporary 2D containment problems are frequently solved with ray casting algorithms~\cite{marussig-17-isogeometricreview}, we note that they can largely be divided into two categories.
In the algebraic approach, intersections between rays and arbitrary curves can be computed using typical root-finding techniques in parameter space~\cite{patrikalakis-10-interrogation}, although standard bisection and gradient descent methods can only identify at most a single point of intersection.
In contrast, geometric approaches take place in physical space, decomposing the curve into subcurves until intersections with the ray can be identified~\cite{farin-01-cagd}.
The current state-of-the art method, \textit{\bezier\ clipping}~\cite{sederberg-90-curveintersections, nishita-90-raytracing}, lies at the intersection of these two approaches, where recursive algebraic subdivision is accelerated using the convex hull property of a \bezier\ curve.
However, such methods are known to possibly report incorrect intersections and suffer from inefficiency in the presence of multiple spatially equivalent intersection points~\cite{efremov-05-robustclipping}. 

These issues are worsened by geometric errors present in the shape.
As an example, if the arbitrary ray is by chance extended through a very small gap in the model, then the point must be classified as ``exterior'' despite this being unintuitive and contrary to the intentions of the mesh designer.
Non-manifold edges in a shape are similarly problematic for ray casting methods.
These issues can be somewhat mitigated by extending several rays and taking a consensus, but such approaches impose an additional computational burden while still producing noisy and potentially inconsistent classifications for nearby points~\cite{nooruddin-03-consensusrepair}.
We now show that, in addition to being indifferent to the specific edge cases introduced by ray casting, winding number algorithms can be extended such that they are also robust to more general issues introduced by non-watertight, non-manifold, curved geometry.

\section{Generalized Winding Numbers}
The generalized winding number is an extension of the standard integer-valued winding number to (potentially) non-watertight regions.
While integer winding numbers partition the enclosed regions of a domain, the generalized winding number generates a harmonic scalar field that smoothly degrades in the presence of discontinuities and self-intersections.
This allows for the calculation of robust containment queries in the presence of messy geometry.
As an example, a simple rule for handling the fractional values that result from imperfections in the bounding geometry is to round them to their nearest integer.

Such generalizations of winding numbers have recently shown great utility in geometry processing applications. 
In the context of both (linear) triangle meshes~\cite{Jacobson-13-winding} and point clouds~\cite{Barill-18-soupcloud}, the resulting inside/outside classification can be used to generate volumetric triangle/tetrahedral meshes out of completely unstructured geometric data.
Additionally, the smooth degradation of the winding number field naturally introduces desirable locality properties, which in turn lead to more robust and performant algorithms in, for example, exact mesh booleans~\cite{trettner-22-ember}.
More recently, these particular winding number algorithms have been used in the context of immersogeometric analysis to simulate fluid flow around geometric objects defined by unstructured point clouds~\cite{balu-23-immersogeometric}.
It is speculated in~\citet{gunderman-21-thesis} that one could extend the calculation of generalized winding numbers by utilizing their own novel quadrature schemes for such objects~\cite{gunderman-20-cad}, but preliminary tests of this approach are limited by the accuracy of numerical integration.
Despite this recent surge in use, methods that compute \textit{exact} generalized winding numbers for more general curved geometric objects are, to our knowledge, currently absent from the literature.

Nevertheless, the problem of computing generalized winding numbers for curved shapes has close analogues in the fields of both computer graphics and boundary integral equations.
In the former, a connection can be drawn to diffusion curves, a primitive in vector graphics that also produces a scalar field defined by the solution of a differential equation with respect to boundary curves (albeit with different data prescribed on the curve)~\cite{Jacobson-13-winding}.
These applications have the advantage of being solved on a grid generated by a rasterized domain, meaning that a global solution can be obtained efficiently using a geometric multi-grid method~\cite{orzan-08-diffusioncurves}. 
\citet{sawhney-20-montecarlogeometry} recently introduced a grid-free approach in which random walks and projections onto curved geometry are used to solve certain classes of PDEs, including that which implicitly defines the generalized winding number, focusing on improving speed and locality at the cost of some degree of accuracy and consistency.
Within the context of boundary integral equations, calculating winding numbers is a special case of the more general problem of integrating double layer potentials with a constant density function. 
However, such algorithms are typically applied to closed domains with simpler (although still curved) types of geometry, precluding general use on arbitrary, messy CAD models~\cite{klinteberg-19-accurate}.
Furthermore, our proposed algorithm relies exclusively on geometric principles, eliminating the need to use costly and potentially inaccurate quadrature schemes.

We begin by reviewing the formal description of a generalized winding number in $\mathbb{R}^2$.
Given an oriented curve $\Gamma \subset \mathbb{R}^2$ and query point ${q} \in \mathbb{R}^2 \setminus \Gamma$, 
the winding number $w$ describes the (potentially incomplete) number of times the curve travels counterclockwise around the point. 
In the case where $\Gamma$ is a closed curve, the scalar field $w_\Gamma({q})$ is integer-valued, and induces a partition of $\mathbb{R}^2$ that can be interpreted as containment in the region bounded by $\Gamma$.
In the following discussion, it is notationally convenient to assume that the query point is located at the origin, and that the curve $\Gamma$ has been appropriately translated by $q$.
The scalar field is invariant to such global translations, and so the winding number of $\Gamma$ at $q$, $w_{\Gamma}(q)$, is identical to the winding number at the origin after translating the curve, $w_{\Gamma-q}(0)$.

When the curve is piecewise linear with $\Gamma = \cup_i L_i$, the winding number at $q$ can be computed as the sum of \textit{signed} angles $\theta_i$ subtended by each linear component $L_i$, given by

\begin{align}\label{eqn:wn_polygon}
    w_{\Gamma-q}(0) &:= \frac{1}{2\pi} \sum_i \theta_i.
\end{align}

\begin{figure}[tbp]
  \centering
  \includegraphics[width=\linewidth]{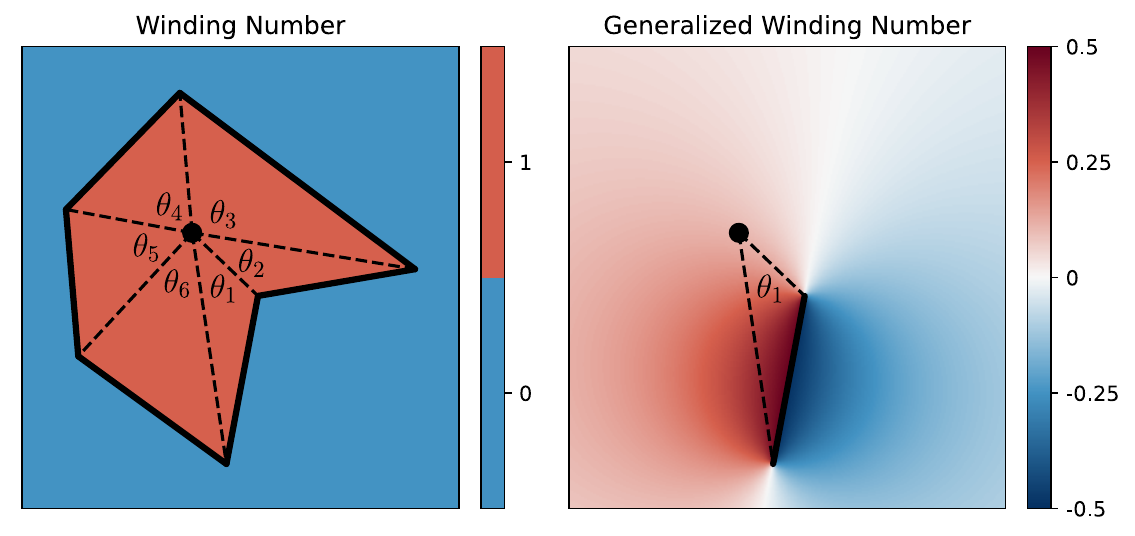}
  \caption{(left) The winding number for a piecewise linear shape can be computed by summing angles subtended by each segment. 
	         (right) Winding numbers for linear segments can be computed at arbitrary points independently of the remaining shape.}
  \Description{Figure 4: The winding number field for a closed polygon is shown to be the sum of the angles subtended by each edge at the point, which is always an integer. The field is then shown for a single edge of the polygon, which is still equal to the angle subtended by the edge at each point in the domain, but no longer an integer anywhere.}
  \label{fig:generalized_winding}
\end{figure}

Furthermore, the winding number of each individual component of the piecewise linear shape (and of straight lines in general) is given by 
\begin{align}\label{eqn:wn_line}
    w_{L_i-q}(0) = \frac{1}{2\pi}\theta_i,
\end{align}
again where $\theta_i$ is the angle subtended by $L_i$ at $q$ (see Figure~\ref{fig:generalized_winding}).

\begin{figure}[tbp]
  \centering
  \includegraphics[width=\linewidth]{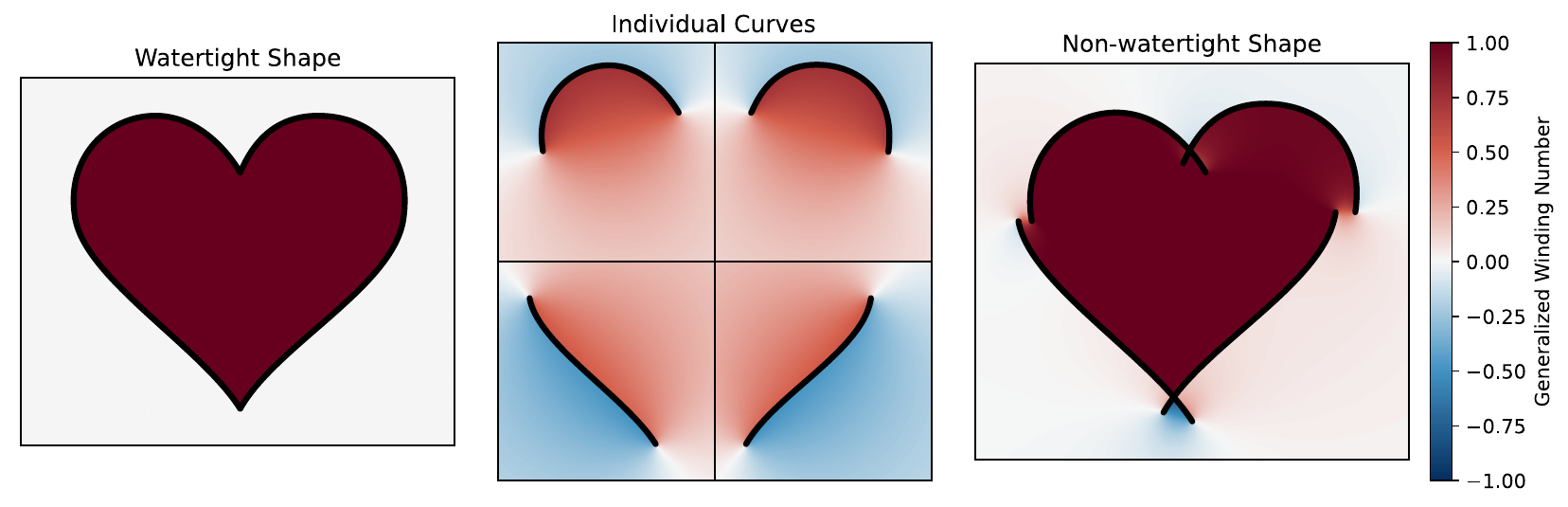}
  \caption{The winding number with respect to closed shapes (left) is always an integer. 
	         However, by independently summing contributions from each curved component (center), 
					the generalized winding number can be computed over collections of curves containing gaps and overlaps (right). }
  \Description{Figure 5: The generalized winding number field is shown for a simple heart shape, which is equal to the integer 1 for all interior points and 0 for all exterior points. The field is then shown for the four curves that compose the shape, which are each computed independently of the others. Finally, the field is shown for the same curves reassembled into a shape that is similar to the original heart, but is not watertight, for which the field is approximately 1 on the interior and approximately 0 on the exterior.}
  \label{fig:generalized_winding_curves}
\end{figure}

We can consider the case for a more general collection of arbitrary curves analogously. 
Let $\Gamma = \cup_i \Gamma_i$ be a collection of rational \bezier\ curves that constitute the (likely not watertight) boundary of a 2D CAD model.
In this case, the winding number of a point $q$ with respect to a single curve $\Gamma_i$ is given by
\begin{align}\label{eqn:wn_polar}
    w_{\Gamma-q}(0) := \frac{1}{2\pi} \int_{\Gamma - q} d\theta = \frac{1}{2\pi}\sum_i\int_{\Gamma_i - q} d\theta
\end{align}
by the properties of the integration.
This is notable, as it describes how the winding number for each individual component $\Gamma_i$ can be computed completely independently of every other curve, with
\begin{align}\label{eqn:wn_curve}
    w_{\Gamma_i-q}(0) := \frac{1}{2\pi} \int_{\Gamma_i-q} d\theta.
\end{align}
It is from this property of the generalized winding number that we derive both the robustness of our winding number calculation, and the robustness of the derived containment query.
While the \textit{collection} of \bezier\ curves may be messy in the sense that pairs of endpoints may not perfectly align, generalized winding numbers on each \textit{individual} curve are well-defined and can be computed stably.
At the same time, the sum of these values at a given point is exactly the generalized winding number of the entire shape, which can be used to produce intuitive and robust containment queries for non-watertight shapes (see Figure~\ref{fig:generalized_winding_curves}).
Most commonly, this is done by rounding this value to the nearest integer, and applying to it a conventional even-odd or non-zero rule to make the classification.
Importantly, each of these conventions treat points with a zero winding number as exterior.

Framed in this way, the remaining task is to evaluate generalized winding numbers for a single, arbitrary rational \bezier\ curve. 
As an initial approach, one could consider direct evaluation of the integral through numerical quadrature. 
When the curve is given parametrically as $(x(t), y(t)) \in \mathbb{R}^2$, $t \in [0, 1]$, as is the case for (rational) \bezier\ curves, we can evaluate this formula more practically in terms of Cartesian coordinates, as

\begin{align}\label{eqn:wn_cartesian}
    w_\Gamma(q) := \frac{1}{2\pi}\int_0^1 \frac{x(t)y'(t) - x'(t)y(t)}{x^2(t) + y^2(t)}\,dt.
\end{align}

\begin{figure}[tbp]
  \centering
  \begin{tabular}{c}
  \includegraphics[width=\linewidth]{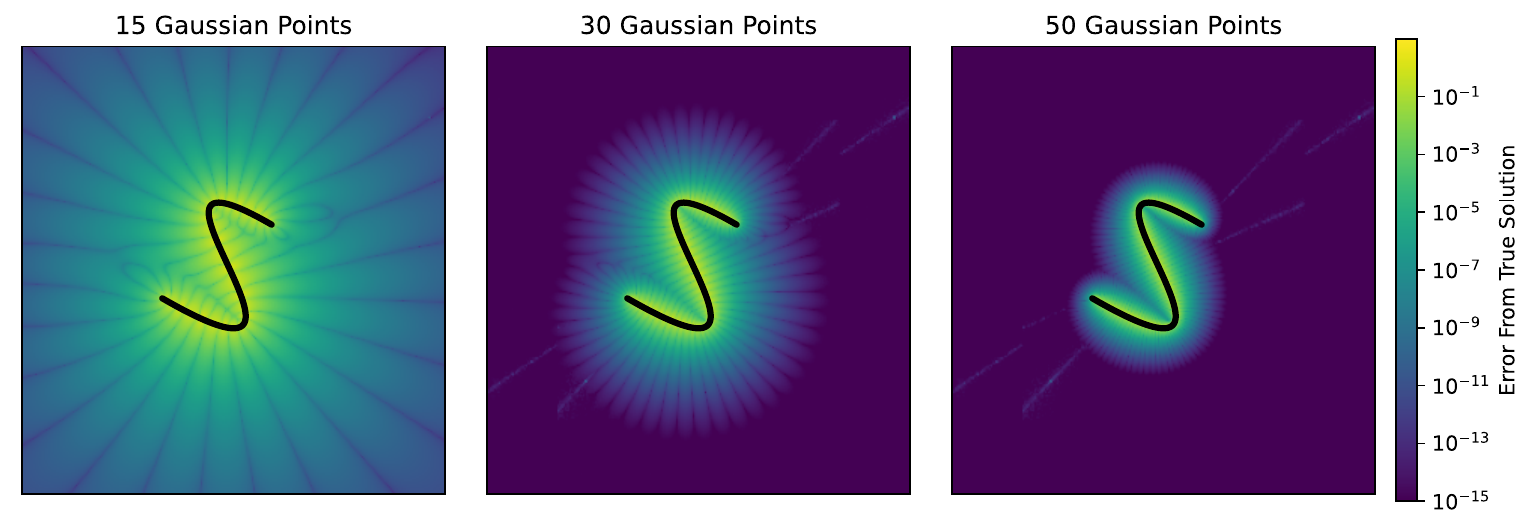} \\
    \includegraphics[width=\linewidth]{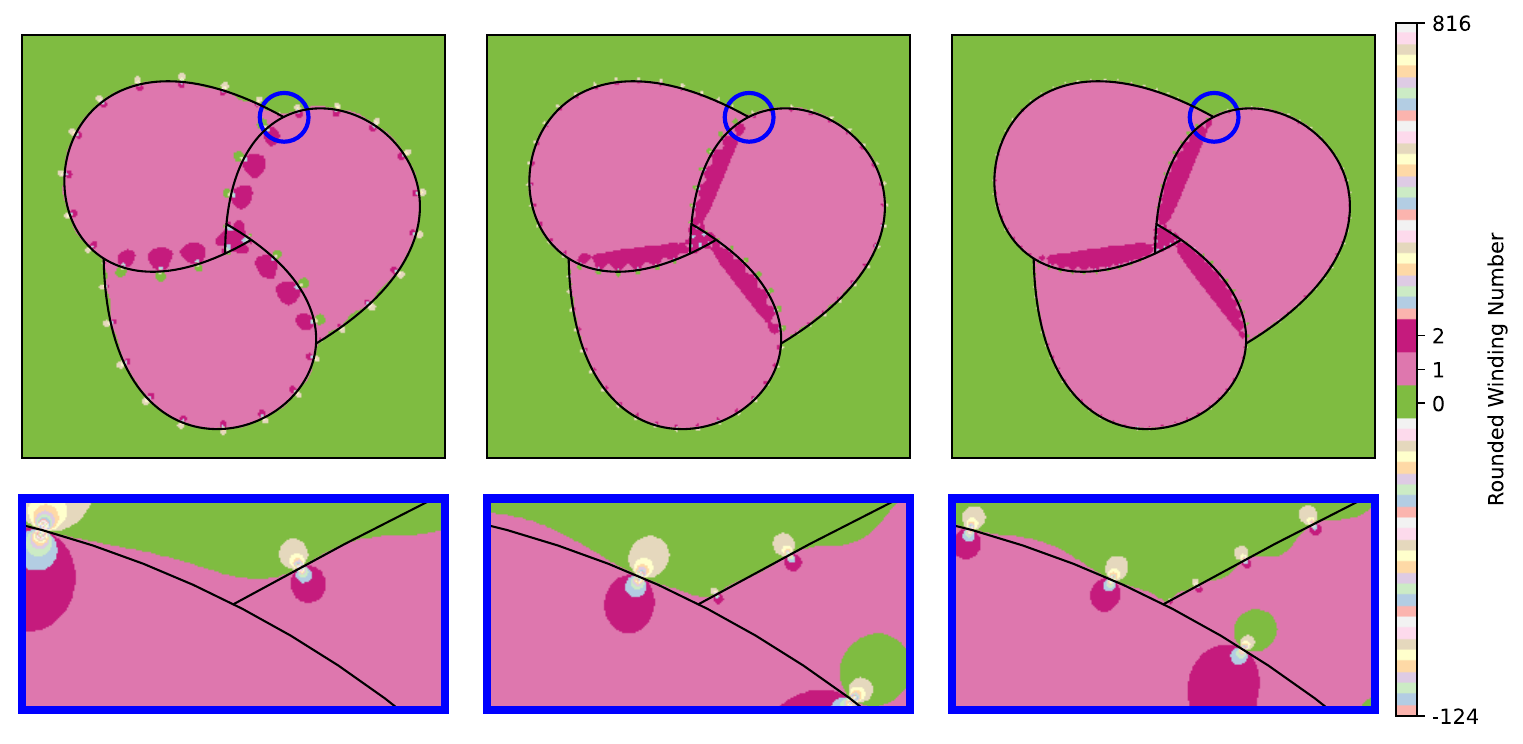}
  \end{tabular}
  \caption{(top) The absolute error (log scale) in Gaussian quadrature used to compute the generalized winding number 
	         on a cubic \bezier\ curve with Equation~\ref{eqn:wn_cartesian}, evaluated with 15-, 30- and 50-nodes. 
					 (middle) Using Guassian quadrature to compute generalized winding numbers over a shape leads to unacceptable errors. 
					(bottom) Close-up of highlighted region.
					}
    \Description{Figure 6: Large numerical error appears when the winding number is computed via quadrature for points near the curve, which is shown on a simple curve. On a simple non-manifold shape, the consequences of this inaccuracy on containment classifications are shown, with even a high-order quadrature rule resulting in many misclassifications.}
\label{fig:quadrature_bad}
\end{figure}

Direct evaluation of this integral through standard techniques is notoriously difficult.
In particular, when $\Gamma$ is close to the query point (or rather the origin under this translated notation) then the near-singular behavior of the integrand causes Gaussian quadrature to become unstable, as seen in Figure~\ref{fig:quadrature_bad}.
Although some error is to be expected for this type of numerical integration, this error rapidly becomes unacceptable as the query point approaches the curve.
Perhaps more concerningly, there is no immediate way to verify that the value returned by the method is accurate.
We can see how this would affect a containment query on a closed shape in Figure~\ref{fig:quadrature_bad}.
In this example, the only correct values for the rounded winding number are 0, 1, and 2.
However, even the most accurate quadrature scheme shown produces values in the vicinity of the quadrature nodes that are completely meaningless.

\section{Generalized Winding Numbers for Curved Geometry}
While these integration formulae are useful from a theoretical perspective, their inherent instability necessitates an approach that avoids evaluating them directly.
In their place, we develop a framework that computes winding numbers based only on geometric properties of the curve.
To this end, we make heavy use of a particular kind of closing curve for each shape. 
Given an (in principle arbitrary) parametric curve $\Gamma$, we define the linear segment $C : [0, 1] \to \mathbb{R}^2$ by $C(t) = {\Gamma(1)}(1- t) + {\Gamma(0)}t$ as the \textit{linear closure} of our curve.

As the name suggests, the union of this closure $C$ and $\Gamma$ will always be a properly oriented, closed curve, and thus partitions $\mathbb{R}^2$ into discrete enclosing regions.
Thus, winding numbers with respect to the total curve $\Gamma \cup C$ are integers, such that
\begin{align}\label{eqn:wnplusclosure}
    w_\Gamma(a) + w_C(a) = w_{\Gamma \cup C}(a) \in \mathbb{Z}.
\end{align}
Most importantly, the winding number of such a closure can always be computed \textit{exactly} without the need to appeal to quadrature by Equation~\ref{eqn:wn_curve}, as the angle subtended by the introduced linear segment can be evaluated through a (relatively) inexpensive arccosine.
In this way, we can always solve the problem of the \textit{generalized} winding number using a solution to the \textit{integer} winding number problem on the closed curve, as subtracting away the contribution of this linear closure can be done independently of the original curve geometry.
While we present our own algorithm to compute this integer winding number in Section~\ref{sec:winding_number_algorithm} that outperforms known alternatives, we note that this strategy is fully compatible with more conventional techniques for computing containment queries in closed regions bounded by curves, e.g.,\ \bezier\ clipping~\cite{nishita-90-raytracing}.

\begin{figure}[tbp]
  \centering
  \includegraphics[width=\linewidth]{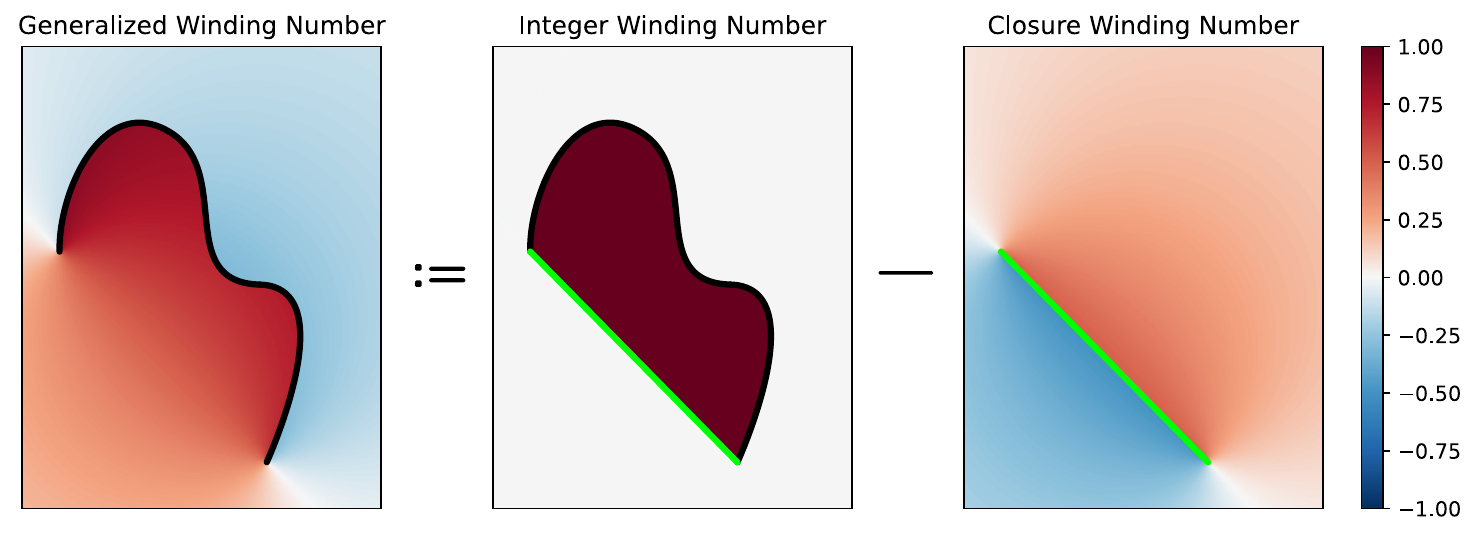}
  \caption{The unknown winding number with respect to a curved shape (left) can be computed by finding the \textit{integer} winding number 
	         of the closed figure (center) and subtracting away the contribution of its closure (right).}
  \Description{Figure 7: A visual demonstration of the generalized winding number is shown for a simple curve. The winding number field for the open curve is unknown, but the field for the closed shape is integer valued, and the field for the closure is known exactly.}
  \label{fig:closure_example}
\end{figure}

This principle is particularly useful in the case when $w_{\Gamma\cup C}(q) = 0$, i.e.,\ the query point is located outside the closed shape $\Gamma\cup C$, or outside the convex hull of $\Gamma$ itself, where we have $w_\Gamma(q) = -w_C(q)$.
This usage is introduced in~\citet{Jacobson-13-winding}, where it is used in a hierarchical evaluation of the winding number for collections of linear triangular facets.
An important consequence of this for curved geometry is that for query points that are far enough away, we can reverse the orientation of $C$ and treat the shape as \textit{equivalent} to its straight line closure, and compute the winding number of the entire curved segment \textit{immediately} and \textit{exactly}.

\begin{figure}[tbp]
  \centering
  \includegraphics[width=\linewidth]{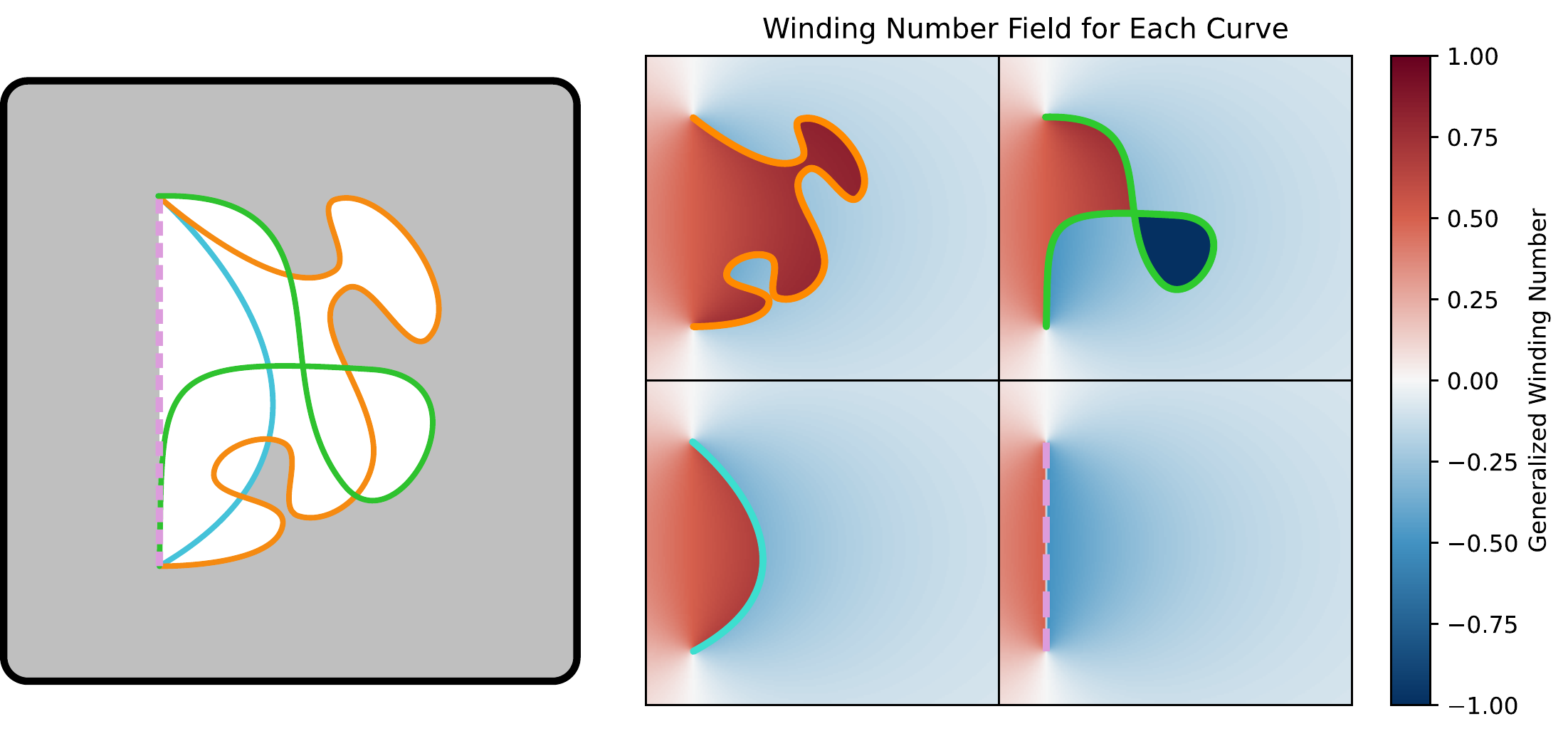}
  \caption{(left) All three curves are closed by the same dashed line, and the shaded region is exterior to all three closed shapes. 
	         (right) This means that the winding number field generated by each curve (and the closure, up to orientation) is \textit{identical} 
					in the shaded region.}
          \Description{Figure 8: A visual demonstration of the principle that the winding number field generated by a curve is closely related to that of its linear closure, shown on four increasingly simple curves which share endpoints.}
  \label{fig:linearize_example}
\end{figure}

We further extend this principle to evaluate generalized winding numbers for points that are arbitrarily close to the curve with the same assurances of exact accuracy.
In such cases, $\Gamma$ can be replaced with a piecewise linear approximation $\widetilde{\Gamma}$ so long as the \textit{integer} winding number at $q$ remains unchanged between $\Gamma\cup C$ and $\widetilde{\Gamma}\cup C$ (see Figure~\ref{fig:linearize_example}).
Doing so necessarily leaves the \textit{generalized} winding number at $q$ unchanged as well, as we have
\begin{align*}
    w_\Gamma(q) = w_{\Gamma \cup C}(q) - w_C(q) = w_{\widetilde{\Gamma} \cup C}(q) - w_C(q) = w_{\widetilde{\Gamma}}(q).
\end{align*} 

It is from these observations that we derive an algorithm for computing \textit{exact} generalized winding numbers over a collection of curves.
For both far and near query points, we construct appropriate linearizations of our curves that are guaranteed to have the same generalized winding number at the point of interest as their curved counterparts.
For a given point, the vast majority of curves in the model will be considered far, and their contribution to the generalized winding number can be computed with a single arccosine evaluation.
For the few curves that are close enough such that the linear closure itself is an insufficient approximation, we construct an approximating polyline that provably generates the same winding number.
In such cases, we evaluate the integer winding number for the closed polygon, which can be done without reference to any trigonometric functions using the point-in-polygon algorithm in~\citet{hormann-2001-pointinpolygon} and subtract away the contribution of the linear closure.

In either case, we are able to compute the \textit{exact} generalized winding number for an arbitrary point using only a single evaluation of arccosine for each curve in the shape.
All that remains is to ensure that this polyline is adaptively constructed such that evaluating the integer winding number of the polyline is efficient.

\section{Winding Number Algorithms}\label{sec:winding_number_algorithm}
We now describe our complete algorithm for solving the generalized winding number problem for a rational \bezier\ curve at a given query point (Algorithm~\ref{alg:windingnumber}). 
The core of this algorithm is described in Algorithm~\ref{alg:integerwindingnumber}, which computes the \textit{integer} winding number for a closed rational \bezier\ curve without reference to ray casting.

In brief, given such a curve $\Gamma$, we adaptively construct a polyline approximate $\widetilde{\Gamma}$ which has the same winding number at the point of interest.
We then apply a standard point-in-polygon algorithm to this closed polyline to compute their shared integer winding number and subtract the contribution of their shared closure.
	
For this procedure to work, we require that our polyline $\widetilde{\Gamma}$ satisfies $w_{\Gamma\cup C}(q) = w_{\widetilde{\Gamma}\cup C}(q)$. 
The simplest way to ensure this is for $q$ to be located outside both closed shapes $\Gamma\cup C$ and $\widetilde{\Gamma}\cup C$, as then this shared integer winding number is equal to 0.
Because this cannot be guaranteed in general, we recursively bisect the curve into components $\{\Gamma_i\}_i$ until $q$ is located outside the closed shape for each.

\begin{figure}[tbp]
  \centering
  \includegraphics[width=0.9\linewidth]{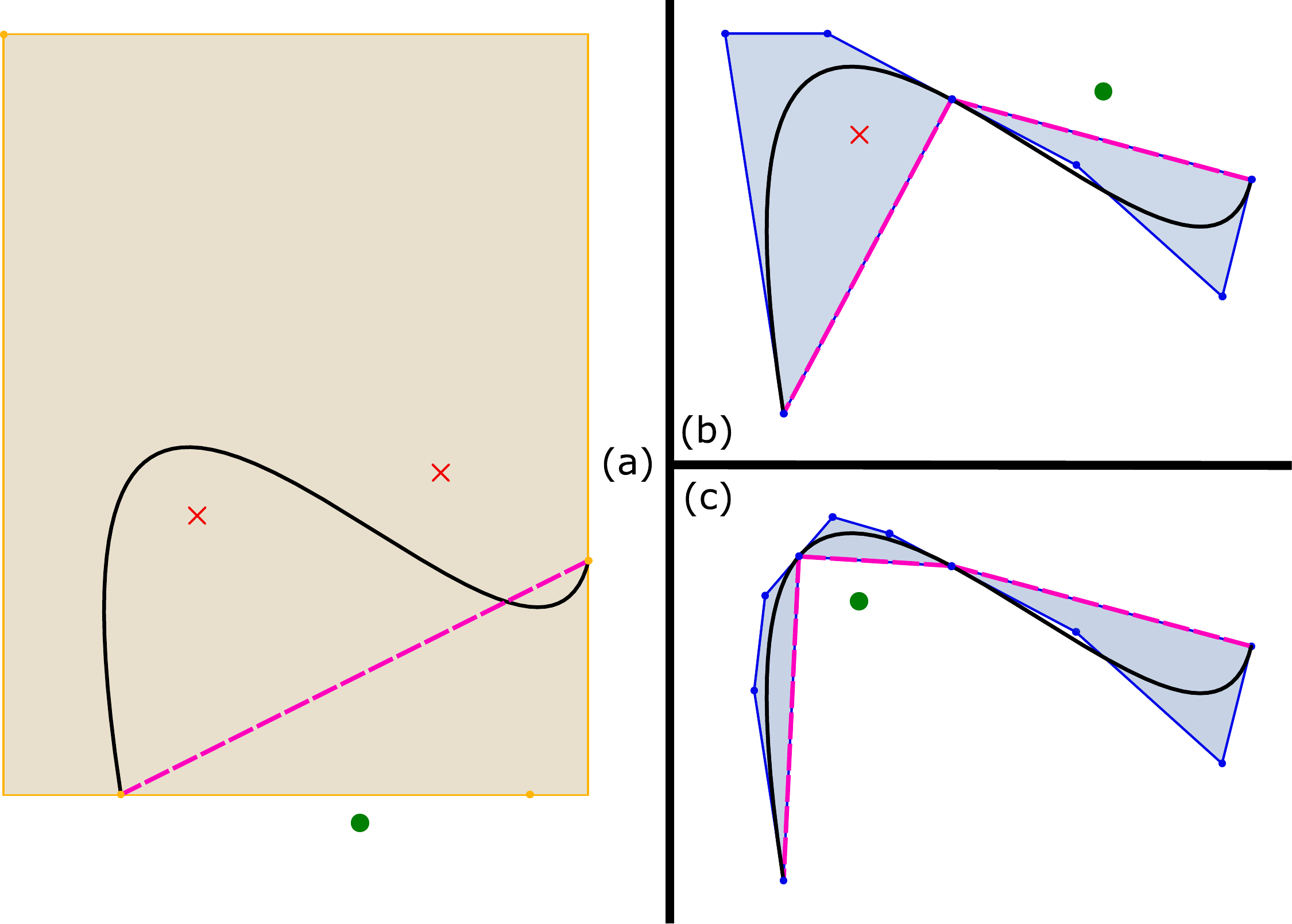}
  \caption{Three iterations of the approximating polyline algorithm. 
      In (a), one point is outside the bounding box, and its winding number can be computed from the dashed closure. 
      In (b), after a bisection we can compute the winding number for an additional point. 
      We repeat the process in (c) to compute the winding number for the remaining point.}
    \Description{Figure 9: Fully described in the caption.}
  \label{fig:polyline_example}
\end{figure}

There are a number of ways to ensure that for each component, $w_{\Gamma_i\cup C_i}(q) = 0$.
For example the convex hull property of a \bezier\ curve states that the curve $\Gamma_i$ is completely contained within the convex hull of its control nodes.
Given a description of this convex hull, classic point-in-polygon algorithms such as~\citet{hormann-2001-pointinpolygon} can be applied to test for containment. 
This leads to the following outline for our algorithm:
\begin{itemize}
    \item If the query point is located outside this convex hull, then it is guaranteed to be outside the closed shape $\Gamma_i \cup C_i$
    and we replace the component $\Gamma_i$ with the reversal of its linear closure $C_i$, as they now provably have the same generalized winding number.
    \item Otherwise, we bisect the \bezier\ curve and repeat the algorithm on each half. As a base case, we check whether the curve is approximately linear (Algorithm~\ref{alg:isapproximatelylinear}) as such segments can be added directly to the polyline.
\end{itemize}
Once this approximating polyline is fully constructed, we close it and use the same point-in-polygon algorithm, \texttt{PolygonWindingNumber}, to compute the integer winding number for the closed polygon.
The contribution of this closure is then subtracted, and we are left with the exact generalized winding number for the curve.
Figure~\ref{fig:polyline_example} illustrates this algorithm for three successively closer query points to a \bezier\ curve.

To improve computational efficiency in the case where subdivision is expensive and exact geometric accuracy is not needed, we optionally provide in Algorithm~\ref{alg:isapproximatelylinear} an early stopping criterion.
We choose this particular heuristic because it is easier to evaluate than the distance between the query point and the curve, and has a more intuitive spatial interpretation than a simple bound on the maximum number of curve subdivisions. 
In practice, \bezier\ curves are smooth nearly everywhere on their interior, and so the recursive bisection step needs to be done relatively few times before the query point is found to be outside the convex hulls of each component, even for points very close to the curve.
Importantly, this causes the accuracy of Algorithm~\ref{alg:windingnumber} and runtime of the procedure to both be reasonably insensitive to this numerical tolerance, as the algorithm is likely to terminate for these smooth curves well before there has been sufficient refinement for subdivisions to be considered linear.
Indeed, in the numerical examples we provide, we set this value to machine epsilon and find no significant loss in performance (See Section~\ref{sec:performance}).

We note that it can be problematic if the query point is located directly on the curve, as in such cases the winding number is not formally defined.
Since we need our algorithm to be compatible to such input points, we discuss this case extensively in Section~\ref{sec:coincident}.

Despite the simplicity of the above procedure, directly computing containment within a \bezier\ curve's convex hull can be prohibitively expensive.
Instead, we first determine if the query is contained in an axis-aligned bounding box that encompasses the convex hull. 
Although this can have a much larger area, the bounding box containment query is inexpensive and most points will be exterior to it.
If it is not, we use a test from~\citet{Ma-03-projection} to check if the control polygon (defined by a \bezier\ curve's control nodes) 
is already simple and convex (see Algorithm~\ref{alg:isvalid}).
If so, we perform the point-in-polygon test to determine if the query lies outside the convex hull of the control polygon.
For efficiency, our implementation notes that when a curve is simple and convex, its subcomponent curves will be as well.

\begin{algorithm}[tbp]
    \caption{\texttt{WindingNumberCurve}: Evaluate the generalized winding number for an arbitrary rational parametric \bezier\ curve}\label{alg:windingnumber}
    \DontPrintSemicolon
    \SetNoFillComment
    \KwIn{$\Gamma$: Rational parametric \bezier\ curve $\Gamma$}
    \myinput{$q$: Query point}
    \KwOut{$w_\Gamma$: The winding number evaluated at ${q}$}
    \nonl\;
    \tcc{Store the linear closure of $\Gamma$}
    $C(t) \gets (1 - t)\Gamma(1) + t\Gamma(0)$\;
    $w_C \gets (1 / 2\pi) \times \Big(\text{Signed angle subtended by } \overrightarrow{qC(0)} \text{ and } \overrightarrow{qC(1)}\Big)$\;
    \eIf{ $q \notin \texttt{BoundingBox}(\Gamma)$}
    {
	    \Return $-w_C$
	}{
		\Return \texttt{IntegerWindingNumberCurve}$(\Gamma \cup C, {q}) - w_C({q})$
    }
\end{algorithm}

\begin{algorithm}
    \caption{\texttt{IntegerWindingNumberCurve} Evaluate the integer winding number for a rational parametric \bezier\ curve closed by a linear segment.
		   We use an algorithm from~\cite{hormann-2001-pointinpolygon} for \texttt{PolygonWindingNumber}.}\label{alg:integerwindingnumber}
    \DontPrintSemicolon
    \KwIn{$\Gamma \cup C$: Closed rational parametric \bezier\ curve}
    \myinput{$q$: Query point}
    \KwOut{$w_{\Gamma \cup C}$The integer winding number evaluated at ${q}$}
    \nonl\;
    $\widetilde{\Gamma} \gets \{ \}$ \tcp*{Initialize the polyline approximation}
    $w_{\Gamma\cup C} = 0$ \tcp*{Initialize the winding number}
    \nonl\;
    Push $\Gamma$ onto an empty stack.\;
    \While{the stack is not empty}{
        $\Gamma_0 \gets \texttt{StackPop}$\;
        $C_0(t) \gets (1-t){\Gamma_0(0)} + t{\Gamma_0(1)}$\;
        \nonl\;
        \tcc{Check for coincidence at the endpoints}
        \eIf{$\texttt{isSimpleConvex}(\Gamma_0)$ \textnormal{\textbf{and}} $(q = \Gamma_0(0)$ \textnormal{\textbf{or}} $q = \Gamma_0(1))$}{
            \tcc{Track the contribution of coincident points}
            $w_{\Gamma \cup C} \mathrel{+}=  $ \texttt{ConvexEndpointWindingNumber}$(\Gamma_0, q)$\;
            $\widetilde{\Gamma} \gets \widetilde{\Gamma} \cup \{C_0\}$\tcp*{Add to the polyline}
        }{
            \eIf{ $\{\texttt{isSimpleConvex}(\Gamma_0)$ \textnormal{\textbf{and}} $q \notin \texttt{ControlPolygon}(\Gamma_0)\}$\\
            \textnormal{\textbf{or}} $\texttt{isApproximatelyLinear}(C_0)$\\}{
             $\widetilde{\Gamma} \gets \widetilde{\Gamma} \cup \{C_0\}$\tcp*{Add to the polyline}
            }{
                $\Gamma_1, \Gamma_2 \gets \texttt{Bisection}(\Gamma_0)$\;
                $\texttt{StackPush}(\Gamma_1, \Gamma_2)$\;
            }
        }
    }
    \nonl\;
    $P \gets \widetilde{\Gamma} \cup \{C\}\;$\tcp*{Close the polyline, forming a polygon}
    \Return $ \texttt{PolygonWindingNumber}(P, q) + w_{\Gamma\cup C}$
\end{algorithm}

\begin{algorithm}[tbp]
    \caption{\texttt{isApproximatelyLinear}: Return \texttt{true} if each control point of a rational parametric \bezier\ curve is within a given tolerance of the closure.}\label{alg:isapproximatelylinear}
    \KwIn{$\Gamma$: Rational parametric \bezier\ curve }
    \DontPrintSemicolon
    \myinput{$\epsilon$: User tolerance}
    \nonl\;
    $C_0(t) \gets (1-t){\Gamma_0(0)} + t{\Gamma_1(1)}$\;
    \For{ each control node ${P}_i$ of $\Gamma$ }{
        \If{$\texttt{SquaredDistance}(C_0, {P}_i) \geq \epsilon$}{
            \Return \texttt{false}\;
        }
    }
    \Return \texttt{true}\;
\end{algorithm}

\begin{algorithm}[tbp]
    \caption{\texttt{isSimpleConvex}: Return \texttt{true} if the polygon of control points of a rational parametric \bezier\ curve is simple and convex (Adapted from~\cite{Ma-03-projection})}\label{alg:isvalid}
    \KwIn{$\Gamma$: Rational parametric \bezier\ curve with control nodes ${P_0}, \cdots, {P_n}$}
    \SetNoFillComment
    \DontPrintSemicolon
    \nonl \;
    \For{ $i = 1, \dots,  (n-1)$}{
        \tcc{Store the linear segment connecting the nodes}
        $S(t) \gets (1 - t) {P_{i-1}} + t{P_{i+1}}$\;
        \eIf{ $i \leq n/2$ }{
            \If{${P_i}$ and ${P_n}$ are on the same side of $S$}{
                \Return \texttt{false}\;
            }
        }{
            \If{${P_i}$ and ${P_0}$ are on the same side of $S$}{
                \Return \texttt{false}\;
            }
        }
    }
    \Return \texttt{true}\;
\end{algorithm}

Altogether, the proposed algorithm is given by Algorithm \ref{alg:windingnumber}.
We note that this procedure is not meaningfully restricted to rational \bezier\ curves.
Their focus throughout this work reflects their simplicity during calculation through manipulation of their control nodes, 
which itself justifies their ubiquity in application. 
More generally, the algorithm is applicable to any curve for which it is possible to construct a bounding box and a linear closure.
In particular, extension to collections of NURBS curves is straightforward.
\section{Generalized Winding Numbers for Coincident Points}\label{sec:coincident}
The generalized winding number is, in a strict mathematical sense, undefined for points located directly on the curve.
This poses a practical problem during the implementation of a containment query, as they are often executed massively in parallel for large clusters of points without any \textit{a priori} knowledge of their position relative to the boundary.
Furthermore, our desire for exact geometric fidelity precludes us from applying the standard trick of perturbing such points randomly to place them definitively on one side of the singular boundary.
In spite of this, it is necessary for compatibility with respect to downstream applications that the algorithm definitively return a value that adheres to a convention that is mathematically justified and intuitive to the caller of \texttt{WindingNumberCurve}.

An important aspect of the winding number scalar field is that it is harmonic, being the unique solution of a Laplacian operator with boundary data prescribed by each side of our curves, enforcing a jump discontinuity across them.
Therefore, the simplest convention for the winding number of coincident points is to take it as the average value across this jump discontinuity.
For linear segments, this boundary data enforces that the winding number approaches $+1/2$ from one side and $-1/2$ on the other.
This means that the winding number should be exactly $0$ at every coincident point along linear segments.
Similarly, this convention would ensure that the winding number along a more general closed shape is a fixed half-integer value along its length, changing only across self-intersections.
However, this presents a unique problem for open, curved shapes, as the winding number is no longer constant along the length of the curve.
This underscores the importance of a single convention for coincident points, as the winding number for individual curves must be computed without any knowledge of the other components that make up the shape, which can be unintuitive if that shape is itself closed.
Nevertheless, we can evaluate the winding number at a coincident point for a single curve just as before, by computing the half-integer winding number of the closed shape and subtracting the contribution of the closing line.

This convention for a coincident winding number is well-justified in a mathematical context as well.
As explained by~\citet{Jacobson-13-winding}, the generalized winding number can be understood in the context of ray casting as the \textit{average} number of intersections from rays cast in all directions from $q$.
When $q$ is on a straight line, only two of the uncountably infinite possible directions will intersect the line, resulting in an average of zero intersections.
Furthermore, the discontinuity in Equation~\ref{eqn:wn_cartesian} is, in fact, removable, and evaluating the remaining integrand agrees with this interpretation of coincident winding numbers.

\begin{figure}[tbp]
  \centering
  \includegraphics[width=\linewidth]{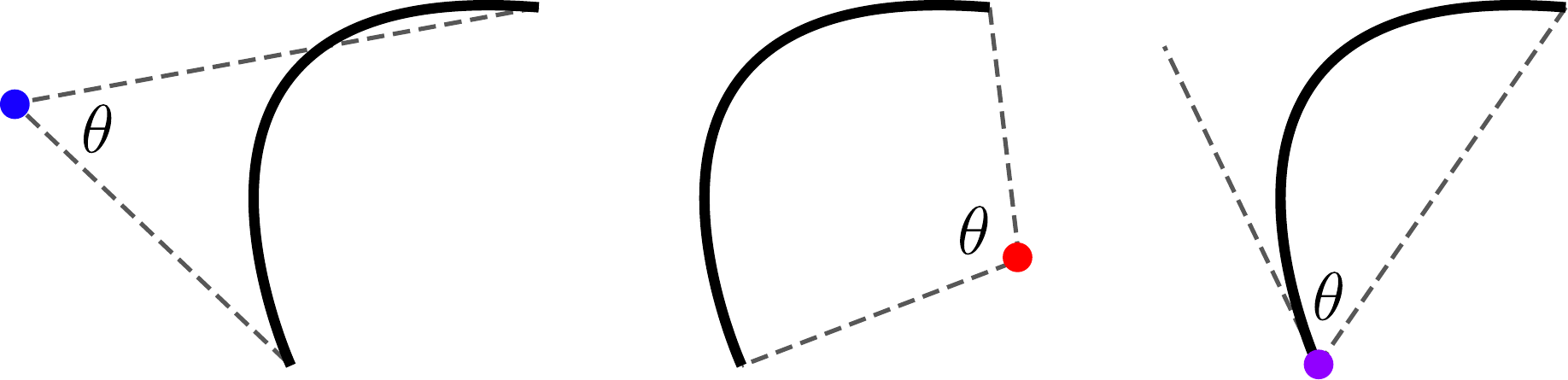}
  \caption{For points located outside the closed curve (left, center), the winding number is proportional 
	         to the angle subtended by the endpoints. For points located directly on an endpoint (right), 
					 we define the winding number as proportional to the angle subtended between the non-coincident endpoint and a tangent vector.}
    \Description{Figure 10: Fully described by the caption.}
  \label{fig:convex_endpoint_example}
\end{figure}

Despite this conceptual clarity, evaluating coincident winding numbers is complicated in practice, as it is computationally expensive to perform the projection necessary to identify when a point is located exactly on a \bezier\ curve~\cite{Ma-03-projection}.
The exception to this is the endpoints, which are interpolated exactly by the first and last control points of the curve.
At such points, the coincident winding number is equal to the signed angle spanned by the non-coincident endpoint, and a \textit{tangent} vector at the coincident endpoint (see Figure~\ref{fig:convex_endpoint_example}).
Furthermore, computing such a tangent is trivial at the endpoint of a \bezier\ curve, as it is defined by the endpoint and the adjacent control node.
Interior points become endpoints in linear time by the repeated bisections of our algorithm.
As this spanned angle must account for full revolutions of this span around the tangent vector, our algorithm only applies this edge-case to curves with simple and convex control polygons (see Algorithm~\ref{alg:convexendpointwindingnumber}).
This particular criterion is typically met after very few bisections, and ensures that no full revolutions can occur on the local subcurve.

\begin{algorithm}[tbp]
    \caption{\texttt{ConvexEndpointWindingNumber}: Return the convention for ``coincident winding numbers'' as outlined in Section~\ref{sec:coincident}}\label{alg:convexendpointwindingnumber}
    \KwIn{$\Gamma$: Rational parametric \bezier\ curve}
    \myinput{$q$: Query point}
    \KwOut{$w_\Gamma$: The winding number evaluated at the endpoint of the curve}
    \DontPrintSemicolon
    \nonl\;
    \eIf{$q = \Gamma(0)$}{
        \Return $\tfrac{1}{2\pi} \times \Big($ Signed angle subtended by $\overrightarrow{q\Gamma(1)}$ and $\overrightarrow{\Gamma'(0)}\Big)$.
    }{
        \Return $\tfrac{1}{2\pi} \times \Big($ Signed angle subtended by $\overrightarrow{q\Gamma(0)}$ and $-\overrightarrow{\Gamma'(1)}\Big)$.
    }
\end{algorithm}
\section{Numerical Experiments and Results}
\subsection{Robustness of Containment Queries on Curved Geometry}

\begin{figure}[tbp]
\centering
  \includegraphics[width=\linewidth]{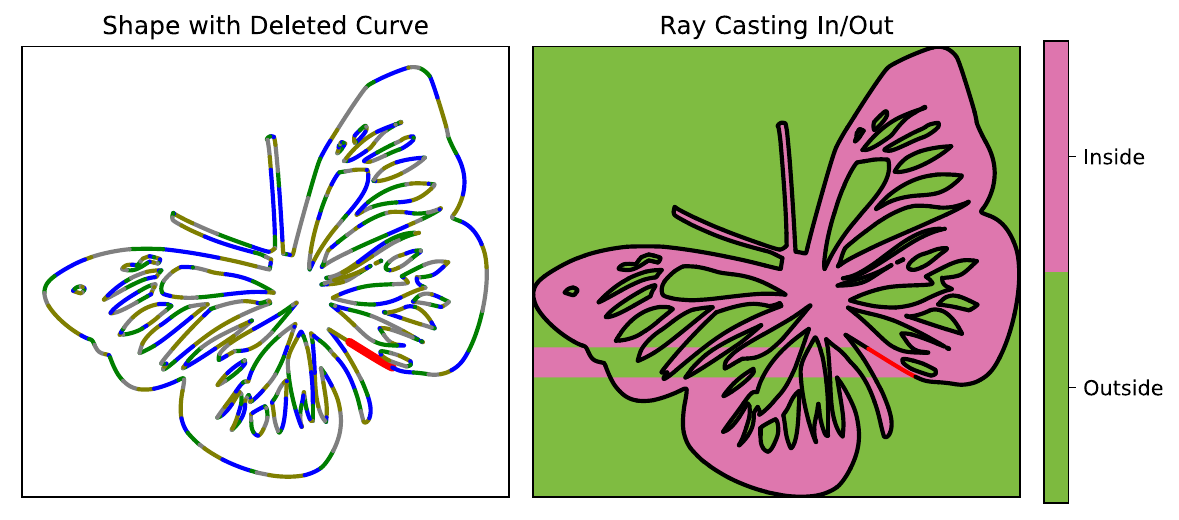}
\caption{(left) A watertight geometric shape, with the exception of the curve in red that is removed during calculation. (right) Ray casting classifies points as interior (pink) and exterior (green). As expected, the geometric error leads to numerous errors in containment queries.}
\Description{Figure 11: A butterfly vector graphic made up of 500+ curves is shown. When one curve is deleted, the ray casting algorithm produces far-reaching errors in containment queries.}
\label{fig:ray_casting_bad}
\end{figure}

We first show the utility of a generalized winding number approach to containment queries in the context of messy CAD geometry.
We compare our approach to conventional ray casting for containment, which remains the de-facto standard for exact containment in curved watertight geometry.
Consider the geometry of Figure~\ref{fig:ray_casting_bad}, which is composed of 87 linear segments and 477 cubic \bezier\ curves.
We remove the marked curve in Figure~\ref{fig:ray_casting_bad} and determine containment at each pixel in the image using a simple ray casting algorithm that extends the ray to the right of the pixel and counts the number of intersections with the shape.
As expected, the use of a ray casting algorithm causes catastrophic issues, in the sense that the containment errors are located at a great distance from the actual geometric error. 
This makes the model unusable in the applications of interest without specific oversight and (often user-driven) correction of these errors.

\begin{figure}[tbp]
    \centering
  \includegraphics[width=\linewidth]{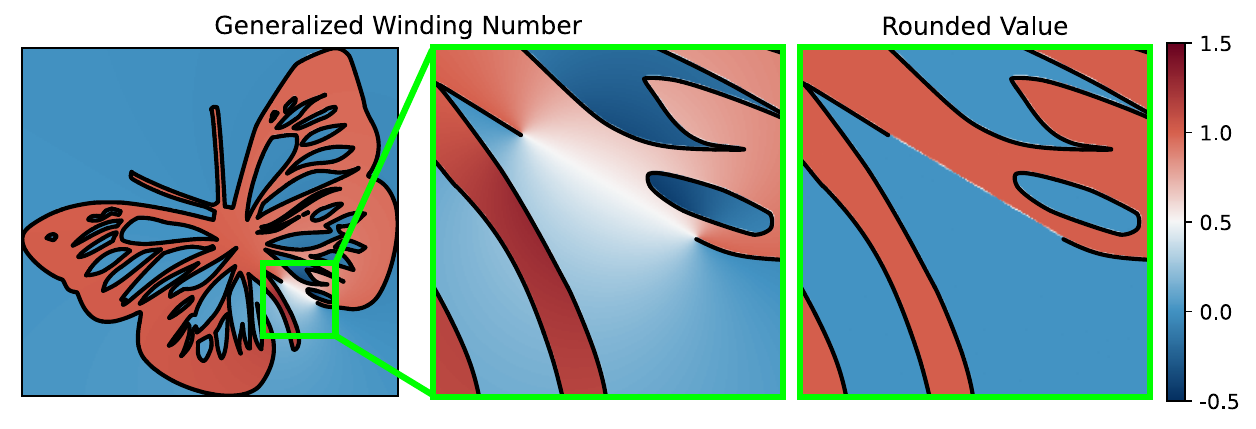}
    \caption{(left) The generalized winding number computed using Algorithm~\ref{alg:windingnumber}. 
            (center) The winding number in the region around the deleted curve degrades smoothly. 
            (right) Rounding the winding number produces an approximation that better conforms to the designer's intuition.}
            \Description{Figure 12: The same butterfly graphic from Figure 11 is shown, but the generalized winding number field is used to determine containment instead. The field degrades smoothly around the deleted curve, and rounding the winding number produces an approximated boundary between interior and exterior regions.}
    \label{fig:winding_number_good}
\end{figure}

We then apply our generalized winding number algorithm to the same geometry in Figure~\ref{fig:winding_number_good}, evaluating at each pixel the winding number field generated by the bounding geometry.
In this example, points which we expect to be interior to this shape have winding numbers close to 1, while the value for points we expect to be exterior is close to 0.
We see that deleting the curve does degrade the field, but that the most severe effects are \textit{localized} to the site of the geometric error.
In some sense, the locality of this degradation is unintuitive, as containment is necessarily a \textit{global} property of points relative to bounding geometry, i.e.,\ the winding number at every point is dependent on every curve.
Importantly, however, the influence of distant curves becomes rapidly negligible.
While the exact value of the winding number now a non-integer value (nearly) everywhere in the domain, away from the deleted curve the difference from the nearest integer is nowhere large enough to have an adverse effect on the actual determination of containment.

A natural mapping of winding numbers to containment classifications is to round each value to the nearest integer and apply a non-zero rule.
For example, in Figure~\ref{fig:winding_number_good} we see that the \textit{rounded} winding number produces a clear boundary between regions along the 1/2 isocurve.
While this implied boundary lacks the curvature of the deleted curve, it is a reasonable approximation and nevertheless allows for the surface to be used immediately in applications.
This behavior is especially desirable when geometric errors arise from small gaps between components of the mesh, as the resulting containment query reflects the simplest closure of the shape permitted by the provided geometric data.

\begin{figure}[tbp]
    \centering
  \includegraphics[width=0.8\linewidth]{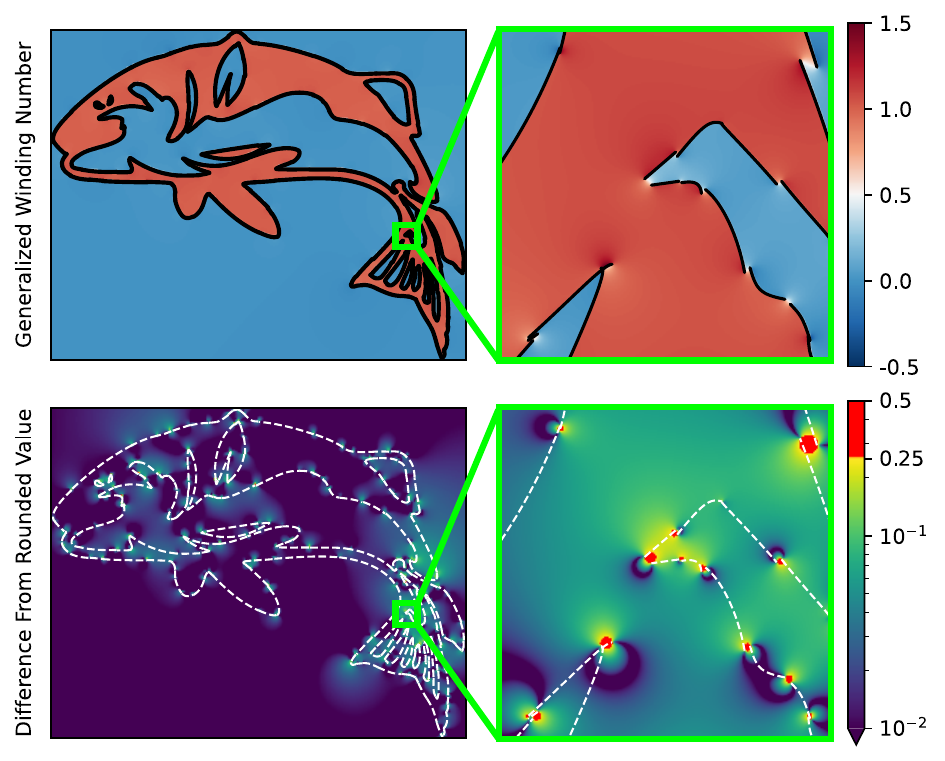}
    \caption{(top) Generalized winding number for a shape that is intentionally deformed to slightly separate adjacent curves. 
             (bottom) Absolute difference (log scale) between computed winding number and rounded winding number. Points for which the difference is greater than 0.25 are highlighted, as this indicates some degree of uncertainty in the classification. 
                       As we see, such points are very sparsely distributed throughout the domain.}
    \Description{Figure 13: A very-slightly non-watertight fish shape is shown with the generalized winding number field computed for each pixel. The absolute difference in log scale between the computed winding number and the rounded winding number is shown, with points that have a difference greater than 0.25 highlighted. These points occur only in small radii around gaps in the shape.}
    \label{fig:fish_figure}
\end{figure}

Through Figure~\ref{fig:fish_figure}, we reiterate the two important kinds of robustness in this work.
The shape in this example is intentionally deformed such that no pairs of adjacent edges are connected.
From a distance, i.e., the scale at which the model geometry is visually ``good enough,'' there are functionally no irregularities in the shape's appearance.
This mirrors the realistic setting, in which such imperceptible tolerances are unexpected and remain catastrophic for conventional containment algorithms.
As before we apply our algorithm to each pixel in the bounding region, and plot the winding number field alongside its difference from the nearest integer.

The first type of robustness is derived from our specific computation of the generalized winding number through Algorithm~\ref{alg:windingnumber}, which is completely robust to non-watertight geometry. 
Because we only ever consider individual curves without any reference to their overall arrangement, both the computational performance and accuracy of our calculation is completely unaffected by the non-watertightness of the shape.
Furthermore, our adaptive algorithm ensures that the winding number is computed stably at arbitrary distances from individual curves.
On the other hand, \textit{any} containment query derived from generalized winding numbers is robust to geometric artifacts in the boundary.
This is because although even small errors have far-reaching influence on the scalar field of winding numbers, this influence degrades rapidly and gracefully away from the sources of these errors.
Even at this frequency of topological errors in this example, the resulting values of the field still very clearly partition the shape into the expected interior and exterior.

It is noted in~\citet{Jacobson-13-winding} that a useful perspective on the fractional value of the winding number is as a measure of confidence for the derived containment query.
For example, points for which the winding number is close to an integer can be considered ``more likely'' to be classified accurately according to the unknown, but presumably watertight shape that lacks any of the present boundary artifacts.
On the other hand, points with winding numbers near half-integer values are often close to the boundaries that are only implied by the rounding heuristic, and correlate with containment classifications that are less informed by existing surface elements.
In Figure~\ref{fig:fish_figure}, we specifically identify each point in the image with a winding number in the range [0.25, 0.75], which are the only points that we can therefore consider to have an ``uncertain'' classification according to our rounding heuristic.
From this, we can see that the \textit{vast} majority of points in this image are classified with high confidence.

Of course, the rounding heuristic is not the only way one can determine containment from a fractional winding number, as access to the full winding number field through our algorithm permits treatment of containment through standard segmentation strategies.
For example, \citet{Jacobson-13-winding} meshes the interior of the shape, utilizing an energy minimization technique to enforce additional smoothness in the resulting segmentation.
Additionally, the above usage of the winding numbers as a measure of confidence can be made more rigorous through further statistical analysis.
In~\citet{sellan-22-stochastic}, the generalized winding number field is used as a prior for a number of statistical distributions, including likelihood of containment.
This rigor is desirable for certain applications, as the fractional value of the winding number alone cannot be interpreted as a probability for the accuracy of the derived containment query.
Because we are ultimately concerned with efficient evaluation of the winding number in this work, we use only the rounding heuristic to determine containment in our examples.

\subsection{Algorithm Performance}\label{sec:performance}

We now consider the computational performance of Algorithms~\ref{alg:windingnumber} and~\ref{alg:integerwindingnumber}.
In each of the following examples, we evaluate our algorithm in the context where exact geometric fidelity is prioritized, and set all numerical tolerances to zero.
For comparison, we consider state-of-the-art techniques that have been adapted to collections of parametric curves.
Perhaps the most common winding number-based approach for watertight geometry would be to discretize the shape into approximating polygons with linear edges and to apply a standard point-in-polygon algorithm to the polygonal shapes to determine containment.
If watertightness cannot be guaranteed, the generalized winding number approaches of~\citet{Jacobson-13-winding,trettner-22-ember} admit a straightforward adaptation to a linearized shape. 

\begin{figure}[tbp]
  \centering
  \includegraphics[width=\linewidth]{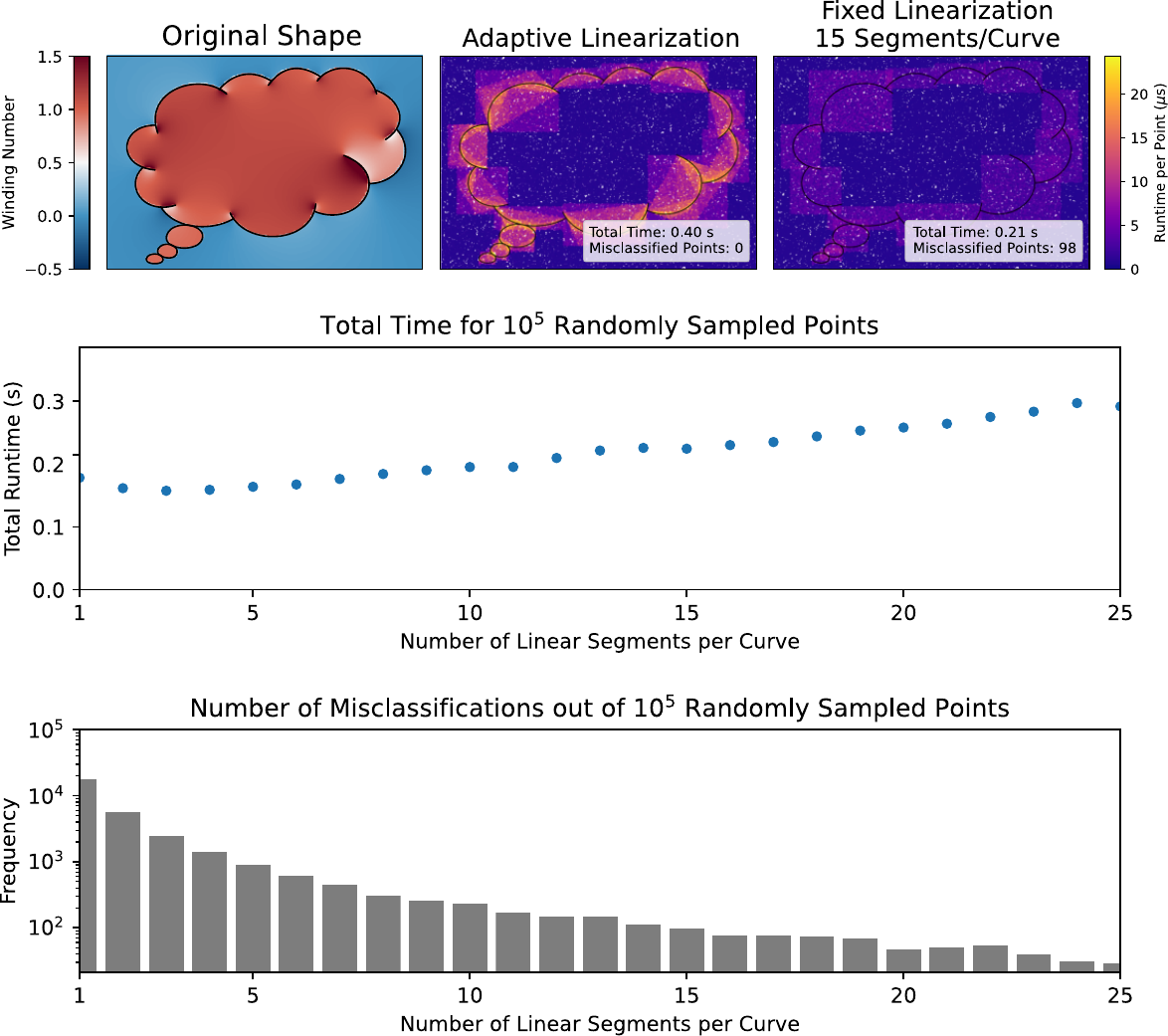}
    \caption{Given a non-watertight, non-manifold shape, we observe the practical effect of curve linearization on geometric accuracy and runtime. We compare our adaptive strategy to those that uses a uniform, piecewise linearization of the curve at fixed levels of refinement. We record the total time spent to evaluate the generalized winding number for $10^5$ points randomly sampled from a uniform distribution, and the number of these points which are ultimately misclassified. We see that a fixed linearization can offer better computational performance, as subdivisions of the curve can be efficiently precomputed. However, we note that even at high levels of refinement there are still a considerable number of misclassifications, which is unacceptable in many downstream applications of interest.}
    \Description{Figure 14: A per-point comparison of the performance of the adaptive strategy to a fixed linearization method is shown. The total wall-clock time needed to evaluate the generalized winding number for all points using various levels of a fixed linearization is shown, along with the number of misclassifications induced by each. The adaptive strategy is slower, particularly for points that are near the curve. However, every level of linearization used introduces a non-zero number of misclassifications.}
  \label{fig:timing_test}
\end{figure}

In demonstration of this fact, consider Figure~\ref{fig:timing_test}, where we compare the performance of our serial implementation of the proposed adaptive strategy to such a linearized method.
Specifically, we consider refinements of each curves in the shape into a fixed number of linear segments.
We note that the shape considered is non-watertight and non-manifold, making it a particularly challenging case for conventional containment queries.
Nevertheless, the fractional winding number field clearly partitions the space according to an intuitive understanding of its interior.

By comparing wall-clock runtime, we see that the use of a fixed level of refinement of the curve leads to a more efficient calculation of the generalized winding number as expected, as curve subdivisions can be computed during a preprocessing step and reused.
While the analogous overhead in the adaptive algorithm can be similarly mitigated through hash maps that cache the results of curve subdivisions, we consider further implementation optimizations for collections of curves to be future work, and instead focus here on efficient and accurate calculation for individual curves.
Indeed, when geometric fidelity is not a concern, using a fixed linearization to compute generalized winding numbers with Algorithm~\ref{alg:windingnumber} is a reasonable approach, although it may involve a considerable memory footprint for larger models.

Importantly, such a linearized method offers no guarantees for geometric fidelity. 
By comparing the level of refinement against an approximate likelihood that points are ultimately misclassified, we see that although the rate of misclassification decreases with increased linear refinement, it does so at a rate that we consider to be far too slow to be practical for many applications of interest, especially for downstream applications that are sensitive to \textit{any} misclassifications.
This is also an issue for methods that sample the bounding geometry non-deterministically, such as those of \citet{nooruddin-03-consensusrepair, Barill-18-soupcloud, sawhney-20-montecarlogeometry}.
In contrast, the proposed method is able to achieve perfect geometric fidelity for points that are arbitrarily close to the boundary. 
In place of these fixed linearization methods, we turn our attention to those that operate directly on curved geometry without approximation.

There remains the issue that the methods known to the authors that operate directly on curved geometry à la \bezier\ clipping~\cite{nishita-90-raytracing} can do so only in the watertight case.
This regrettably means that no direct comparison for the performance Algorithm~\ref{alg:windingnumber} can be made at this time.
In its place, we note that our procedure for computing the exact generalized winding number for any open, curved object requires applying an \textit{arbitrary} integer winding number algorithm on the closed shape and subsequently subtracting out the contribution of the closure of the curve.
It is in this context that we demonstrate the performance of Algorithm~\ref{alg:integerwindingnumber}, which is one such algorithm that computes the integer winding number of a closed, curved object.
We show that Algorithm~\ref{alg:integerwindingnumber} is considerably more performant than alternatives utilizing crossing numbers for containment queries on closed, curved geometry, as such methods necessarily introduce inefficiencies through a reliance on ray casting.

We first consider a common procedure for ray casting that applies a geometric \textit{binary search} to compute ray-\bezier\ curve intersections~\cite{farin-01-cagd}.
In this algorithm, we extend from the query a ray in the direction of the closest edge of a bounding box, and recursively bisect the curve until guarantees for potential intersections can be made.
If a subcurve is sufficiently linear (see Algorithm~\ref{alg:isapproximatelylinear}) such that an intersection is guaranteed, then the signed intersection can be recorded.
If the ray does not intersect the subcurve's bounding box, then it can be discarded.
This procedure converges to points of intersection linearly, and is capable of identifying multiple intersections with the same curve.

\bezier\ clipping~\cite{nishita-90-raytracing,sederberg-90-curveintersections} is a similar procedure, discarding sections of the curve that are guaranteed to have no intersections with the ray.
However, the convergence of this approach to points of intersection is quadratic, as the curve is instead split into three subcurves at each iteration.

To make the appropriate comparison to Algorithm~\ref{alg:integerwindingnumber} in the context of exact winding numbers for non-watertight geometry, we utilize each \textit{within} the framework described in Algorithm~\ref{alg:windingnumber}. 
That is to say, we iterate over each open curve in the model and close it, compute the integer winding number with Algorithm~\ref{alg:integerwindingnumber} or one of the above alternate techniques, and subtract out the contribution of the closure.
We note that, while these alternate techniques themselves are well-known in the context of ray-tracing and containment of closed shapes, computing generalized winding numbers in this way is, to our knowledge, itself a novel application of their use.

In comparing these three algorithms, there are a number of specific implementation details that govern the wall-time of their evaluation, in particular the use of a spatial index to efficiently handle far away points.
However, we are concerned with their efficiency in the near-curve regime, where the number of \textit{curve evaluations} (most commonly occurring as a result of a curve bisection) is an effective proxy for computational efficiency between the methods.
Furthermore, the computation of the exact fractional value for the generalized winding number only involves the evaluation of a single arccosine for each curve in the model, and this cost is equivalent across all three methods.

\begin{figure}[tbp]
  \centering
  \includegraphics[width=\linewidth]{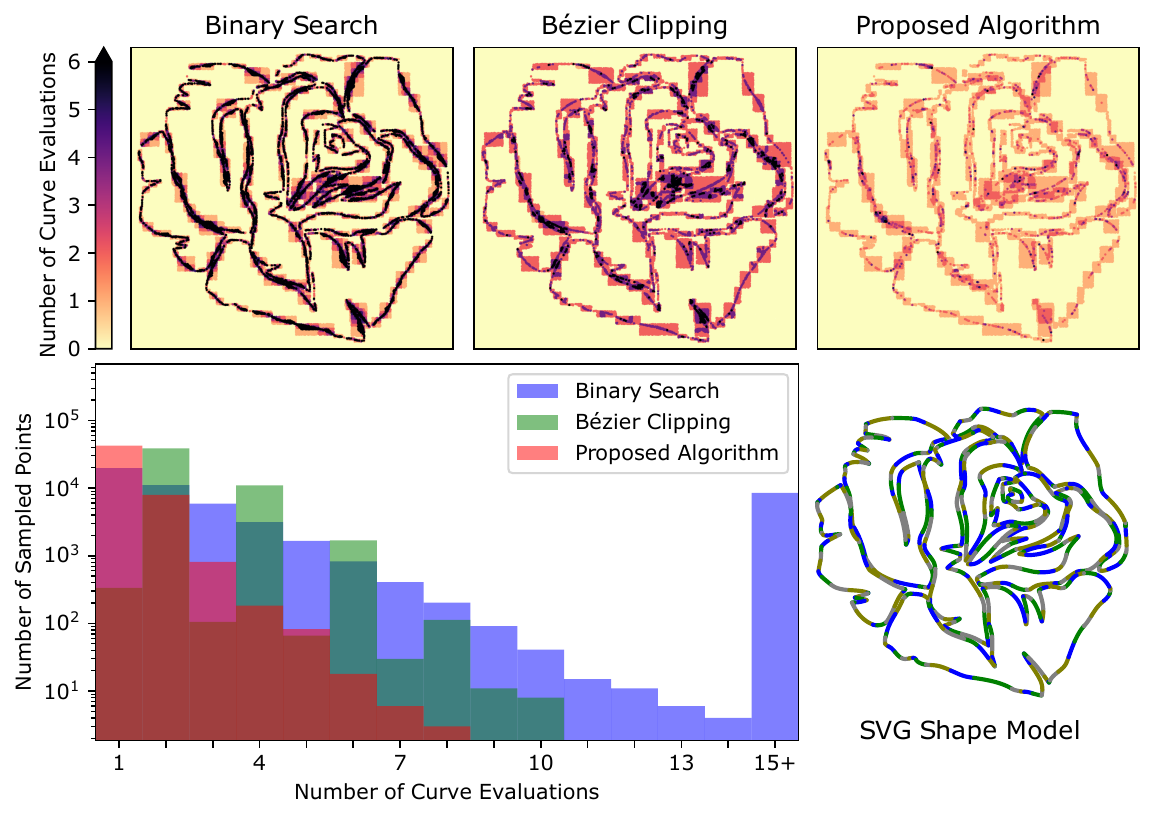}
  \caption{We compare the cost of a conventional geometric bisection method, \bezier\ clipping, and our proposed method on a watertight model. 
    We uniformly sample 250,000 points from a bounding box and measure the number CAD model evaluations, i.e., curve subdivisions, 
	  that are necessary to evaluate the winding number at each point to full geometric precision. 
	  In the histogram, we omit the number of sampled points that require zero function evaluations, 
	  as this occurs whenever the point is outside each curve's bounding box, occurring at the same frequency for each method.}
    \Description{Figure 15: Given a watertight rose shape, a heatmap of the number of curve evaluations needed to evaluate the generalized winding number for 250,000 points in the domain is shown for each of a binary search method, Bezier clipping, and the proposed method. The proposed method requires the fewest curve evaluations for the vast majority of points. A histogram shows the distribution of curve evaluations needed for each method.}
  \label{fig:rose_figure}
\end{figure}

Our first example in Figure~\ref{fig:rose_figure} uses a relatively simple shape featuring at most polynomial, cubic \bezier\ curves.
To compare the three algorithms (binary search, \bezier\ clipping, and the proposed approach presented in Algorithm~\ref{alg:integerwindingnumber}), we randomly sample 250,000 points from a bounding box around the shape using a uniform distribution, and count the number of curve evaluations on each that needs to be made to determine containment with perfect geometric fidelity. 
Again, we emphasize that all three methods are being used here to compute the generalized winding number, which they do by computing the integer winding number of the closed shape for each curve in the model.

As expected, the binary search ray casting algorithm performs the poorest, with a considerable number of points requiring more than 15 curve subdivisions to make an evaluation. 
This is because the terminating condition of the naive algorithm is linearity of the curve component near the intersection, for which the number of bisections increases with curvature.
This behavior is improved considerably with \bezier\ clipping, which performs more sophisticated subdivision (at greater computational expense) by considering the convex hull of the curve. 
However, the proposed algorithm improves on this further, with no points requiring more than 8 function evaluations, and most requiring only one or two.

Furthermore, we recognize that for the vast majority of points, \textit{zero} function evaluations are required for each algorithm.
This is because the first containment check performed in each is one to an axis-aligned bounding box for the entire curve, and most points lie outside the bounding box for most curves.
For this reason, integer winding numbers for most combinations of points and curves are known to be zero without any curve evaluations.

\begin{figure}[tbp]
  \centering
  \includegraphics[width=\linewidth]{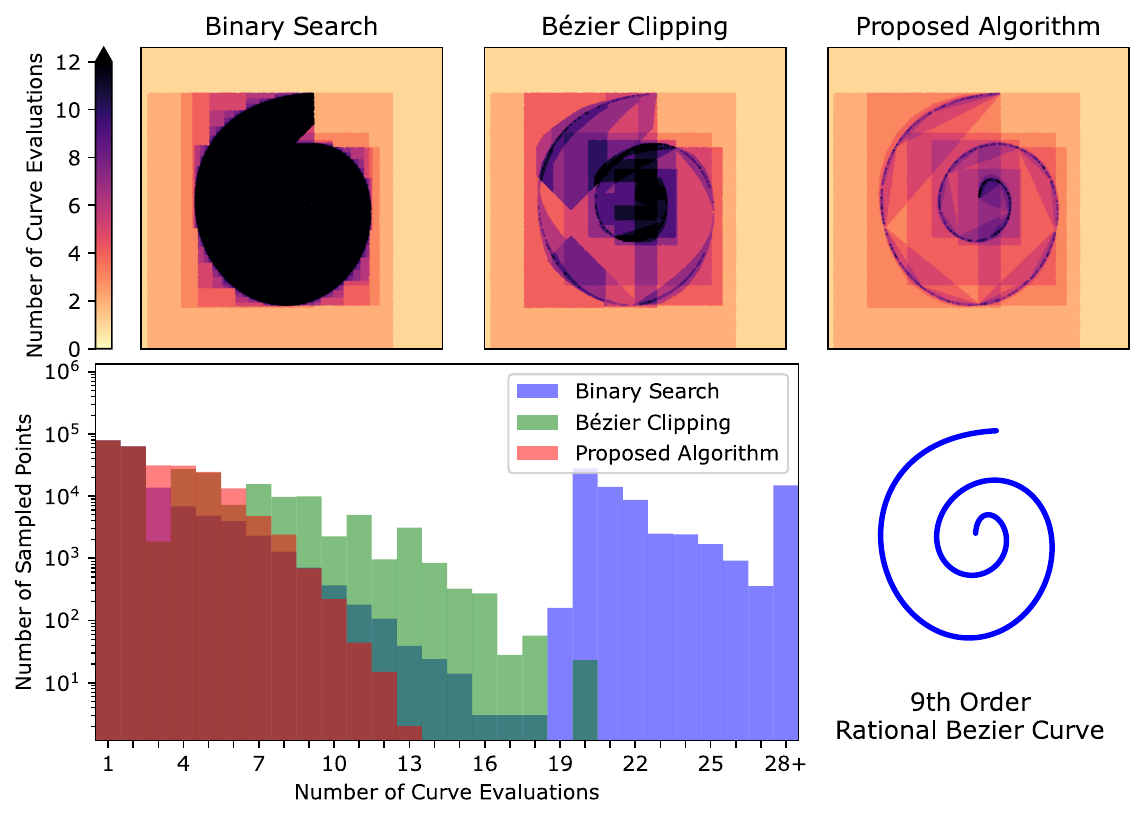}
  \caption{For higher-order curves, additional curve subdivisions are necessary to reach perfect geometric fidelity. 
	         The proposed algorithm still has superior performance in such cases, 
					 as the location of the various intersections need not be computed directly.}
    \Description{Figure 16: An identical plot to Figure 15 is shown, but on a high-order rational Bezier curve in the shape of a spiral. While all three methods need more evaluations than for the simpler watertight shape, the proposed method still requires the fewest curve evaluations for the vast majority of points.}
  \label{fig:rational_figures}
\end{figure}

One explanation for the relatively poor performance of the two ray casting algorithms is that such procedures are unnecessarily informative.
Not only do ray casting algorithms \textit{identify} potential intersections, but they also inherently \textit{locate} the intersections, information which is rarely needed when the ray itself is chosen arbitrarily.
In contrast, by not needing to compute the location of these intersections, the proposed algorithm is able to converge with considerably fewer curve evaluations, despite only doing so linearly.
This is particularly relevant in the case where the curves of interest contain multiple overlapping components, self-intersections, or regions of significant curvature, the latter of these cases being tested directly on a rational curve in Figure~\ref{fig:rational_figures}. 
In these examples, the ray casting algorithms perform particularly poorly, as any ray extended will intersect the curve multiple times, incurring additional cost with each.

\begin{figure}[tbp]
  \centering
  \includegraphics[width=\linewidth]{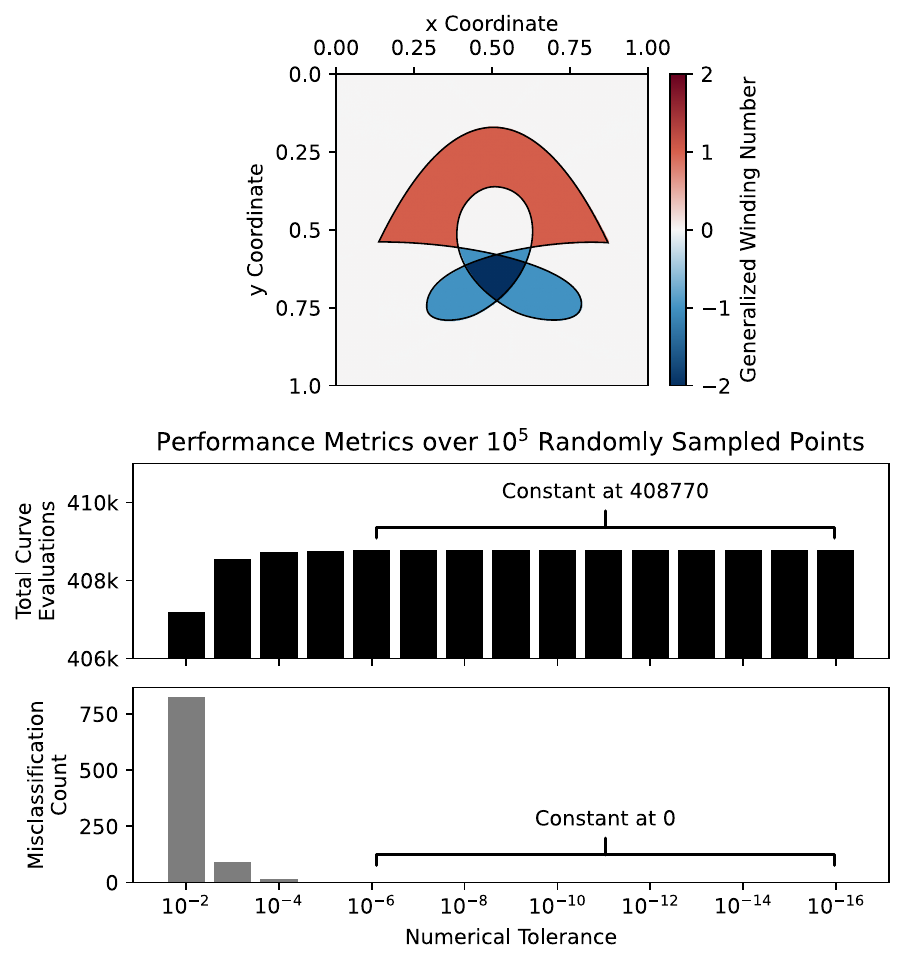} 
  \caption{Beyond a certain threshold, the adaptive nature of the proposed algorithm causes it to be insensitive to the numerical tolerance in Algorithm~\ref{alg:isapproximatelylinear}.
  In practice, we set this tolerance to zero, as this ensures exact geometric fidelity while not incurring a large additional computational burden.}
  \Description{Figure 17: Two bar plots show the number of curve evaluations and misclassifications for the proposed method as a function of the numerical tolerance for the Algorithm 3. These results are reported for randomly sampled points around a high-order rational Bezier curve. When the tolerance is set lower than 10 to the power of -6, the proposed method uses a fixed number of curve evaluations and does not introduce any misclassifications.}
  \label{fig:tolerance_test}
\end{figure}

Finally, we consider the sensitivity of our procedure for computing generalized winding numbers to the numerical tolerance introduced in Algorithm~\ref{alg:isapproximatelylinear}.
We do this by considering performance and accuracy as a function of this tolerance, respectively measured in terms of the number of curve evaluations required to determine containment and the number of misclassifications caused by the final layer of implied linearization.
These metrics are evaluated over $10^5$ points sampled from a bounding box for the closed shape in Figure~\ref{fig:tolerance_test}, which for demonstration purposes depicts a $9^\text{th}$ order, rational curve that far exceeds the complexity of those typically used in practice.
We see that invoking Algorithm~\ref{alg:isapproximatelylinear} to prematurely terminate the procedure does marginally reduce the overall computational cost, but that this performance improvement is closely correlated with losses in classification accuracy.
This effect is amplified on shapes with such complex curvature, as it becomes more likely that the implied linearization imposed by the tolerance will produce misclassifications.
At the same time, we observe that even for very intricate shapes, the performance of Algorithm~\ref{alg:integerwindingnumber} rapidly stabilizes along both metrics.
As a result, our practical preference is to set this tolerance to zero even in cases where geometric accuracy is not a priority.

\begin{figure}[tbp]
  \centering
  \includegraphics[width=0.75\linewidth]{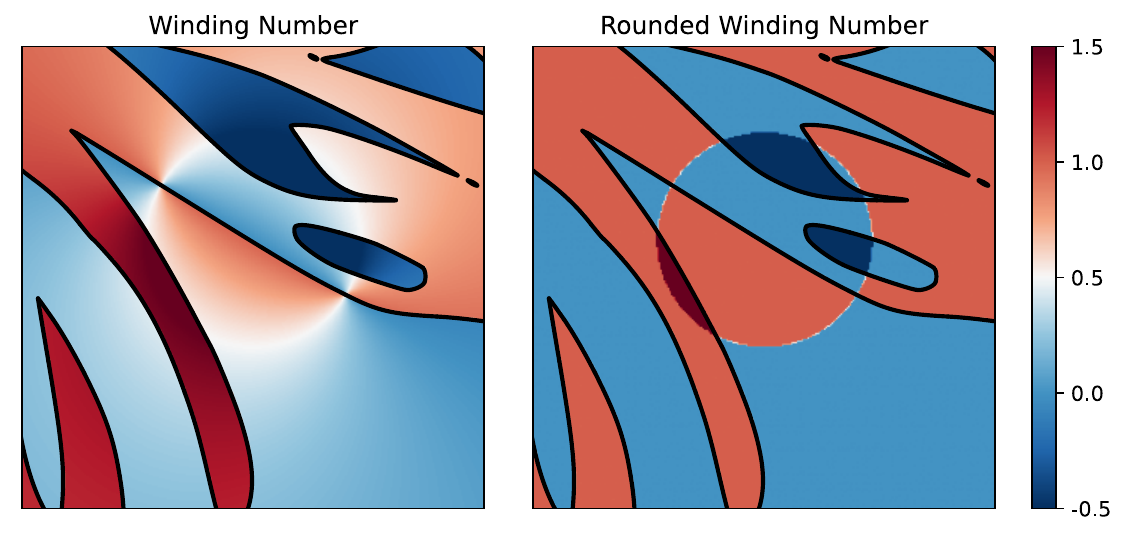}
  \caption{Generalized winding numbers generated by the shape in Figure~\ref{fig:winding_number_good}, 
	   but with the orientation of the highlighted curve reversed rather than removed entirely. 
		 The reversal of this curve causes surrounding containment classifications to be swapped within a local region.}
     \Description{Figure 18: The same butterfly graphic from Figure 11 is shown, but one curve has the orientation reversed instead of deleted. The reversal causes the winding number field to be reversed in a circular region around the curve.}
  \label{fig:orientation_bad}
\end{figure}

As is always the case when generalized winding numbers are used to determine containment, there is still an assumption that the collection of curves are reasonably \textit{oriented}. 
Because the scalar field generated by a single curve is computed completely independently of every other curve by design, a single reversed curve interferes with the contribution of surrounding curves, functionally reversing nearby containment queries, as can be observed in Figure~\ref{fig:orientation_bad}.
While the induced classification errors are still local to the geometric error, and would no doubt cause catastrophic errors \textit{without} the use of generalized winding numbers, they nevertheless represent a type of geometric error unaccounted for by the current method.

\section{Concluding remarks}

In this paper, we proposed a generalized winding number algorithm for collections of parametric curves, 
addressing the inherent challenges for curved geometry that are not shared by their linear counterparts.
For the vast majority of points that are considered ``sufficiently far'' from a given curve, we circumvent the need for expensive linearization of the shape by treating the curve directly as a single segment connecting its endpoints.
For all other points, we demonstrated that our recursive algorithm outperforms existing ray casting algorithms adapted for this context.
Finally, we formalize a procedure for handling winding numbers of points that are coincident with the curve, further increasing the practicality of our algorithm.
In each case, this algorithm makes minimal reference to the often near-singular integrals from which the winding number is theoretically derived, instead operating strictly according to geometric and trigonometric principles.
Nevertheless, we are able to \textit{exactly} compute generalized numbers for curved shapes in two dimensions.
Whereas the focus of this work has been on efficiently and robustly computing winding numbers for individual curves within a collection,
follow-up work will consider accelerating the overall query workflow using a spatial index, as in~\citet{Jacobson-13-winding,Barill-18-soupcloud,weiss-16-shaping} and exploiting the inherent parallelism within Algorithm~\ref{alg:windingnumber} via threaded and/or GPU implementation.

In future work, we are interested in developing a generalized winding number for curved surfaces in $\mathbb{R}^3$, 
but anticipate considerable challenges in developing an efficient numerical implementation.
While much of the mathematical underpinnings remains the same, we cannot merely replace polar angles with their solid angle analogues.
For example, while a line can close any curve in 2D, there are no simple geometric primitives that can close an arbitrary surface in 3D, or at the very least none that have direct formulas for their winding numbers.
Nevertheless, the potential impacts of such a technique is highly motivating in spite of the challenges, and its pursuit will likely have theoretical foundations in the work presented here.

\begin{acks}
We would like to thank the anonymous reviewers for their thoughtful consideration and insightful feedback.
This work performed under the auspices of the U.S. Department of Energy by Lawrence Livermore National Laboratory under Contract DE-AC52-07NA27344. LLNL-TR-828702.
\end{acks}

\bibliographystyle{ACM-Reference-Format}
\bibliography{citations}

\end{document}